\documentclass[10pt,final,twocolumn,twoside]{IEEEtran} 
\usepackage{cite}
\usepackage{amssymb,psfrag,graphicx,subfigure,textcomp,bm}
\usepackage[dvips]{color}
\usepackage{epsf,epsfig,pstricks, pst-plot}
\usepackage{stfloats}
\usepackage[cmex10]{amsmath}
\usepackage{acronym}

\newcommand\Matrix[2]
{   \left[ \begin{array}{#1}
        #2
    \end{array} \right]
}

\newtheorem{defi}{Definition}
\newtheorem{thm}{Theorem}
\newtheorem{cor}{Corollary}

\newtheorem{pro}{Proposition}

\newcommand\E   {  \mathbb{E} }
\newcommand\T   {  \textrm{tr} }

\newcommand\NA  { { \mathcal{N}_\Tx{a} } }
\newcommand\Na  { { N_\Tx{a} }}
\newcommand\NB { { \mathcal{N}_\Tx{b} } }
\newcommand\Nb { { {N}_\Tx{b} } }

\newcommand\V[1]    {  \mathbf{#1} }
\newcommand\B[1]    {  \boldsymbol{#1} }

\newcommand\Tx[1]   {   \textrm{#1} }

\newcommand\J   {  \V{J} }
\newcommand\JE[1]   {  {\V{J}_\Tx{e}({#1})} }

\newcommand\R   {  \V{J}_\Tx{r} }

\newtheorem{remrk}{Remark}

\newcounter{MYtempeqncnt}

\begin{document}


\title{Cooperative Network Navigation: Fundamental Limit and its Geometrical Interpretation}

\author{Yuan~Shen,~\IEEEmembership{Student~Member,~IEEE},
		Santiago~Mazuelas,~\IEEEmembership{Member,~IEEE},
        and Moe~Z.~Win,~\IEEEmembership{Fellow,~IEEE}
\thanks{Manuscript received February 15, 2011; revised July 20, 2011; accepted August 29, 2011. 
This research was supported, in part, by the National Science Foundation under Grant ECCS-0901034,
			the Office of Naval Research under Grant N00014-11-1-0397, 
			and
			the MIT Institute for Soldier Nanotechnologies.}
\thanks{The authors are with the Laboratory for Information and Decision Systems (LIDS), Massachusetts Institute of Technology, 77 Massachusetts Avenue, Cambridge, MA 02139 USA (e-mail: {shenyuan@mit.edu},  {mazuelas@mit.edu}, {moewin@mit.edu}).}
\thanks{Digital Object Identifier XXX.XXX}
}

\maketitle

\markboth{IEEE Journal on Selected Areas of Communications, Vol.~X, No.~Y, Month~2011}{Shen \emph{\MakeLowercase{et al.}}: Cooperative Network Navigation: Fundamental Limit and its Geometrical Interpretation}

\begin{abstract}
Localization and tracking of moving nodes via \emph{network navigation} gives rise to a new paradigm, where nodes exploit both temporal and spatial cooperation to infer their positions based on intra- and inter-node measurements. While such cooperation can significantly improve the performance, it imposes intricate information processing that impedes network design and operation. In this paper, we establish a theoretical framework for cooperative network navigation and determine the fundamental limits of navigation accuracy using equivalent Fisher information analysis. We then introduce the notion of carry-over information, and provide a geometrical interpretation of the navigation information and its evolution in time.
Our framework unifies the navigation information obtained from temporal and spatial cooperation, leading to a deep understanding of information evolution in the network and benefit of cooperation.
\end{abstract}


%
%

\section{Introduction}\label{sec:Intr}

Real-time reliable localization and tracking capability is a key enabler for numerous emerging applications in commercial, public safety, and military sectors. These include logistics, security tracking, medical services, search and rescue operations, vehicle networking, and military operations \cite{SayTarKha:05, PahLiMak:02, CafStu:98, SheWin:J10a, PatAshKypHerMosCor:05,  WymLieWin:J09, SheWymWin:J10, JouDarWin:J08, GezTiaGiaKobMolPooSah:05, WinConMazSheGifDarChi:J11}. This wide range of potential applications has motivated an increasing research interest in localization and tracking technologies for wireless networks \cite{PlaKum:08, PavCosMazConDar:06, KhaKarMou:09, VerDarMazCon:B08, DarConBurVer:07, ConDarDecWin:J12, DarCon:04, RabOppDen:06, YuMonRabCheOpp:06, XuSheMarCai:07, LuoZha:07, Mai:08, DarConFerGioWin:J09, MazLorBah:J10, TicMurNeh:98}. 

Navigation can be thought of as an inference process encompassing both localization and tracking, where mobile nodes (agents) in a network determine their positional states\footnote{The positional state commonly includes the position, velocity, acceleration, orientation, and angular velocity.} based on measurements and prior knowledge. In conventional systems, each agent individually determines its positional state by using its own measurements with respect to fixed infrastructures and/or from inertial sensors. For instance, in the Global Positioning System (GPS), each agent infers its position based on the pseudorange measurements taken with respect to multiple satellites with known positions \cite{Spi:78}; and in self-tracking systems, each agent infers its positional state based on the inertial measurements about its movement \cite{Fox:05}. However, these conventional techniques fail to provide satisfactory performance in many scenarios: GPS-based navigation becomes inaccurate in harsh and/or indoor environments (e.g., in buildings and urban canyons) due to signal blockage, while inertial-based navigation becomes inaccurate in long-term operation due to velocity drift. 

\begin{figure}[t]
	\psfrag{x11}[l][][1.5]{\hspace{-5mm} $\V{x}_{1}^{(1)}$}
	\psfrag{x12}[l][][1.5]{\hspace{-5mm} $\V{x}_{1}^{(2)}$}
	\psfrag{x21}[l][][1.5]{\hspace{-5mm} $\V{x}_{2}^{(1)}$}
	\psfrag{x22}[l][][1.5]{\hspace{-4mm} $\V{x}_{2}^{(2)}$}
	\psfrag{z121}[l][][1.5]{\hspace{-6mm} $\{\V{z}_{12}^{(1)}, \,\V{z}_{21}^{(1)}\}$}
	\psfrag{z122}[l][][1.5]{\hspace{-10mm} $\{\V{z}_{12}^{(2)}, \,\V{z}_{21}^{(2)}\}$}
	\psfrag{z111}[l][][1.5]{\hspace{-6mm} $\V{z}_{11}^{(1)}$}
	\psfrag{z221}[l][][1.5]{\hspace{-5.5mm} $\V{z}_{22}^{(1)}$}
	
	\psfrag{A}[l][][1.8]{\hspace{-5mm} Agent 1}
	\psfrag{B}[l][][1.8]{\hspace{-3mm} Agent 2}
	\psfrag{C}[l][][1.8]{\hspace{-9mm} Agent 3}
	
	\centering
	{\includegraphics[angle=0,width=0.82\linewidth,draft=false]{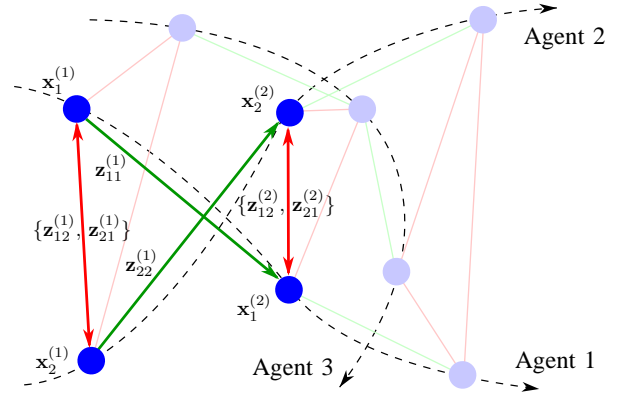}}

    \caption{Network navigation: a network with three agents (blue circles) in three time steps. Intra-node measurements $\{\V{z}_{kk}^{(n)}\}$ and inter-node measurements $\{\V{z}_{kj}^{(n)}\}$ are denoted by green and red arrows, respectively. \label{fig:concept}}
\end{figure}

Cooperative techniques are attracting increasing interest for localization \cite{SheWymWin:J10, WymLieWin:J09, ChaSah:06, Lar:04, SavRabBeu:01, PatAshKypHerMosCor:05}, driven by the success of cooperative techniques in many wireless applications \cite{FreKowKum:10, DohLi:10, BhaHjoSon:08, LiHanPoo:08, HosKimBha:10}. Such techniques have been shown to improve localization performance due to sharing of information among spatial neighbors \cite{SheWymWin:J10,WymLieWin:J09}. In addition to spatial cooperation, the agents in the network can also benefit from information obtained in different time instants (see Fig.~\ref{fig:concept}). This temporal information is traditionally exploited by each agent independently using filtering techniques in a process commonly known as self-tracking. Here we advocate the joint use of temporal and spatial cooperation for real-time localization and tracking in wireless networks, referred to as network navigation.

In the framework of network navigation, agents in a network jointly infer their positional states by sharing information in both the temporal and spatial domains, referred to as \emph{temporal cooperation} and \emph{spatial cooperation}, respectively. Agents obtain information associated with spatial neighbors by inter-node measurements related to nodes' relative  positions (e.g., ranges); as well as from prior knowledge about the spatial environment.\footnote{Examples of inter-node measurement sensors include RF radios, vision sensors, and GPS devices; examples of  prior spatial knowledge include positions of certain nodes and map information.} In addition, agents obtain information associated with temporal evolution by intra-node measurements related to temporal derivatives of positions (e.g., velocities); as well as from prior knowledge about the temporal variation of positions.\footnote{Examples of intra-node measurement sensors include inertial measurement unit, odometer, and compass; examples of prior temporal knowledge include mobility models and the types of moving agents.}


The proper use of all relevant information through cooperation in a navigation network can lead to dramatic performance improvements over conventional approaches. However, realizing the benefits of temporal and spatial cooperation incurs associated costs such as additional communication and complex information fusion. Thus, a deep understanding of the fundamental limits of cooperative navigation networks is important not only for providing performance benchmarks but also for guiding network design and operation under the performance/complexity trade-off. 

In this paper, we establish a general framework for cooperative network navigation to determine the fundamental limits of navigation accuracy by Fisher information analysis. Built on our results of static localization using only spatial cooperation \cite{SheWymWin:J10}, we incorporate intra-node measurements and mobility knowledge as another component of cooperation, i.e., temporal cooperation. The main contributions of the paper are as follows:
\begin{itemize}
	\item We determine the fundamental limits of navigation accuracy, in terms of navigation information, for both Bayesian and deterministic formulations of cooperative network navigation.
	\item We derive the navigation information by using equivalent Fisher information (EFI) analysis, and show that the EFI matrix can be decomposed into basic building blocks corresponding to each measurement and prior knowledge in temporal and spatial cooperation.
	\item We develop a geometrical interpretation of the navigation information, providing insights into information evolution in network navigation. 
\end{itemize}

The rest of the paper is organized as follows. Section \ref{Sec:Model} presents the network setting, the Bayesian and deterministic models, as well as the notion of EFI. In Section \ref{sec:EFIM_Navigation}, we derive the EFI matrices (EFIMs) for network navigation based on both Bayesian and deterministic models. Then in Section \ref{sec:geometrical_interpretation}, we investigate the EFIM for 2-D navigation problem and develop a geometrical interpretation. Finally, conclusions are drawn in the last section.

\subsubsection*{Notation}
The notation $\E_{\V{x}}\{\cdot\}$ is the expectation operator with respect to the random vector $\V{x}$; $\V{A} \succeq \V{B}$ denotes that the matrix $\V{A} - \V{B}$ is positive semi-definite; $\T\{\cdot\}$, $\left[\, \cdot \, \right]^\dagger$, $|\cdot|$, and $\text{adj}(\, \cdot \,)$ denote the trace, transpose, determinant, and adjugate matrix of its argument, respectively; $\|\cdot\|$ denotes the Euclidean norm of its argument; and $\mathbb{S}_{++}^2$ and $\mathbb{S}_{+}^2$ denote the set of $2\times 2$ positive-definite and positive-semidefinite matrices, respectively. We define the unit vectors $\V{u}_{\phi} := [\,\cos\phi\; \sin\phi\,]^\dagger$ and ${\V{u}}_{\phi}^\perp := \V{u}_{\phi+\pi/2}$. The notation $\V{x}_{k_1:k_2}$ is used for concatenating the set of vectors $\left\{ \V{x}_{k_1}, \V{x}_{k_1+1}, \ldots, \V{x}_{k_2}\right\}$ and similarly $\V{x}_{k_1:k_2}^{(t_1:t_2)}$ for ${\big\{}\V{x}_{k_1:k_2}^{(t_1)}, \V{x}_{k_1:k_2}^{(t_1+1)},\ldots, \V{x}_{k_1:k_2}^{(t_2)}{\big\}}$, for $k_1\leq k_2$, $t_1 \leq t_2$. We also denote by $f(\V{x})$ the probability density function (pdf) $f_\V{X}(\V{x})$ of the random vector $\V{X}$ unless  otherwise specified.


\section{Problem Formulation}\label{Sec:Model}

In this section, we describe the network setting, formulate the problem of network navigation, and briefly review the notion of EFI that will be used to characterize the fundamental limits of navigation accuracy.

\subsection{Network Setting}\label{Sec:Model_SysMod}

Consider a cooperative wireless network consisting of $\Na$ agents (or targets) and $\Nb$ anchors (or beacons), where each agent is equipped with both intra- and inter-node measurement sensors. The sets of agents and anchors are denoted by $\NA = \{1,2, \dotsc, \Na \}$ and $\NB = \{\Na+1,\Na+2, \dotsc, \Na+\Nb \}$, respectively. Both the measurements and navigation inference are made in discrete time instants $t_n$'s $(n=1,2,\ldots,T)$. Let $\V{x}_k^{(n)} \in \mathbb{R}^D $ 
be the positional state of node $k$ at time $t_n$, where those of the agents are to be determined. The positional state includes the position $\V{p}_{k}^{(n)}$ and other mobility parameters 
such as velocity, acceleration, orientation, and angular velocity.\footnote{For example, $D=8$ for 2-D navigation in this case.} 

The parameter vector at time $t_n$ is given by 
\begin{align}\label{eq:parameter_vector}
	\B\theta^{(n)} = \Matrix{cccccccc}{\V{x}_{1:\Na}^{(n)\,\dagger} & \B{\eta}_{1:\Na}^{(n) \,\dagger} & \B\kappa_{1:\Na}^{(n) \,\dagger} }^\dagger
\end{align}
where $\B{\eta}_{k}^{(n)}$ and $\B\kappa_{k}^{(n)}$ include the parameters of node $k$ associated with intra- and inter-node measurements, respectively.\footnote{Examples of parameters associated with intra- and inter-node measurements are clock drift and wireless channel parameters, respectively.} 
In particular, $\B\kappa_{k}^{(n)}$ is the concatenation of the set of vectors ${\big\{} \B\kappa_{kj}^{(n)}: \, j\in\NA\cup\NB\setminus\{k\} {\big\}}$ associated with inter-node measurements between node $k$ and other nodes.\footnote{We consider the general case where the pairwise parameters of inter-node measurements are distinct, i.e., $\B\kappa_{kj}^{(n)} \neq \B\kappa_{jk}^{(n)}$. For example, $\B\kappa$ are the channel parameters for nonreciprocal wireless channels \cite{Mol:05, Mol:09}. Our results can be easily specialized to the case when $\B\kappa_{kj}^{(n)} = \B\kappa_{jk}^{(n)}$ by eliminating the duplicate from the parameter set.} Correspondingly, the set of measurements made at time $t_n$ is denoted by $\V{z}^{(n)} = {\big\{}\V{z}^{(n)}_{kj}:k\in\NA,\, j\in\NA\cup\NB{\big\}}$, where $\V{z}_{kk}^{(n)}$ denotes the intra-node measurements made at node $k$, and $\V{z}_{kj}^{(n)}$ denotes the inter-node measurements made at node $k$ with respect to node $j \neq k$.

\subsection{Bayesian and Deterministic Models}

In this subsection, we describe the Bayesian and deterministic models for network navigation.

\subsubsection{Bayesian model}\label{sec:Bayesian_model}

\begin{figure}[t]
	\psfrag{t1}[l][][2.5]{\hspace{-6mm} Time 1}
	\psfrag{t2}[l][][2.5]{\hspace{-6mm} Time 2}
	\psfrag{x3_1}[l][][1.5]{\hspace{-4mm} $\V{x}_3^{(1)}$}
	\psfrag{x1_1}[l][][1.5]{\hspace{-4mm} $\V{x}_1^{(1)}$}
	\psfrag{x2_1}[l][][1.5]{\hspace{-4mm} $\V{x}_2^{(1)}$}
	\psfrag{x3_2}[l][][1.5]{\hspace{-6mm} $\V{x}_3^{(2)}$}
	\psfrag{x1_2}[l][][1.5]{\hspace{-4mm} $\V{x}_1^{(2)}$}
	\psfrag{x2_2}[l][][1.5]{\hspace{-4mm} $\V{x}_2^{(2)}$}

	\psfrag{k12_1}[l][][1.5]{\hspace{-3mm} $\B{\kappa}_{12}^{(1)}$}	
	\psfrag{k12_2}[l][][1.5]{\hspace{-3mm} $\B{\kappa}_{12}^{(2)}$}
	\psfrag{z12_1}[l][][1.5]{\hspace{-3mm} $\V{z}_{12}^{(1)}$}	
	\psfrag{z12_2}[l][][1.5]{\hspace{-3mm} $\V{z}_{12}^{(2)}$}	
	
	\psfrag{k21_1}[l][][1.5]{\hspace{-2mm} $\B{\kappa}_{21}^{(1)}$}	
	\psfrag{k21_2}[l][][1.5]{\hspace{-2mm} $\B{\kappa}_{21}^{(2)}$}
	\psfrag{z21_1}[l][][1.5]{\hspace{-3mm} $\V{z}_{21}^{(1)}$}	
	\psfrag{z21_2}[l][][1.5]{\hspace{-3mm} $\V{z}_{21}^{(2)}$}	

	\psfrag{k13_1}[l][][1.5]{\hspace{-7mm} $\B{\kappa}_{13}^{(1)}$}	
	\psfrag{k13_2}[l][][1.5]{\hspace{-7mm} $\B{\kappa}_{13}^{(2)}$}
	\psfrag{z13_1}[l][][1.5]{\hspace{-4mm} $\V{z}_{13}^{(1)}$}	
	\psfrag{z13_2}[l][][1.5]{\hspace{-4mm} $\V{z}_{13}^{(2)}$}

	\psfrag{k31_1}[l][][1.5]{\hspace{-4mm} $\B{\kappa}_{31}^{(1)}$}	
	\psfrag{k31_2}[l][][1.5]{\hspace{-4mm} $\B{\kappa}_{31}^{(2)}$}
	\psfrag{z31_1}[l][][1.5]{\hspace{-3mm} $\V{z}_{31}^{(1)}$}	
	\psfrag{z31_2}[l][][1.5]{\hspace{-3mm} $\V{z}_{31}^{(2)}$}
	
	\psfrag{k23_1}[l][][1.5]{\hspace{-4mm} $\B{\kappa}_{23}^{(1)}$}	
	\psfrag{k23_2}[l][][1.5]{\hspace{-4mm} $\B{\kappa}_{23}^{(2)}$}
	\psfrag{z23_1}[l][][1.5]{\hspace{-4mm} $\V{z}_{23}^{(1)}$}	
	\psfrag{z23_2}[l][][1.5]{\hspace{-4mm} $\V{z}_{23}^{(2)}$}
	
	\psfrag{k32_1}[l][][1.5]{\hspace{0mm} $\B{\kappa}_{32}^{(1)}$}	
	\psfrag{k32_2}[l][][1.5]{\hspace{0mm} $\B{\kappa}_{32}^{(2)}$}
	\psfrag{z32_1}[l][][1.5]{\hspace{-2mm} $\V{z}_{32}^{(1)}$}	
	\psfrag{z32_2}[l][][1.5]{\hspace{-2mm} $\V{z}_{32}^{(2)}$}
	
	\psfrag{zi1_1}[l][][1.5]{\hspace{-4mm} $\V{z}_{11}^{(1)}$}	
	\psfrag{zi1_2}[l][][1.5]{\hspace{-4mm} $\V{z}_{11}^{(2)}$}
	\psfrag{zi2_1}[l][][1.5]{\hspace{-4mm} $\V{z}_{22}^{(1)}$}	
	\psfrag{zi2_2}[l][][1.5]{\hspace{-4mm} $\V{z}_{22}^{(2)}$}
	\psfrag{zi3_1}[l][][1.5]{\hspace{-5mm} $\V{z}_{33}^{(1)}$}	
	\psfrag{zi3_2}[l][][1.5]{\hspace{-5mm} $\V{z}_{33}^{(2)}$}
	\psfrag{n1_1}[l][][1.5]{\hspace{-3.5mm} $\B\eta_{1}^{(1)}$}
	\psfrag{n1_2}[l][][1.5]{\hspace{-3.5mm} $\B\eta_{1}^{(2)}$}
	\psfrag{n2_1}[l][][1.5]{\hspace{-3.5mm} $\B\eta_{2}^{(1)}$}
	\psfrag{n2_2}[l][][1.5]{\hspace{-3.5mm} $\B\eta_{2}^{(2)}$}
	\psfrag{n3_1}[l][][1.5]{\hspace{-5mm} $\B\eta_{3}^{(1)}$}	
	\psfrag{n3_2}[l][][1.5]{\hspace{-5mm} $\B\eta_{3}^{(2)}$}	
	
	\centering
	{\includegraphics[angle=0,width=0.95\linewidth,draft=false]{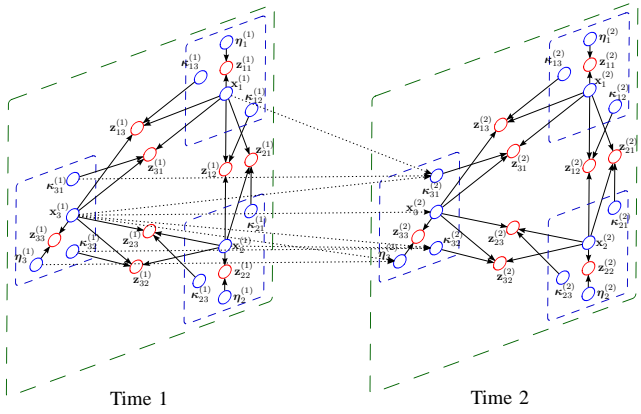}}

    \caption{Graphical model describing the interrelationships among the parameters and measurements present in the Bayesian model for cooperative network navigation. The temporal cooperation is illustrated only for agent 3. \label{fig:Bayesian}}
\end{figure}

The agents' positional states and measurement parameters can be described as random variables. In this case, the evolution in time of such variables and measurements is characterized by a hidden Markov model (HMM),\footnote{HMMs have been widely applied as a statistical modeling tool since they allow modeling complex real-world problems with reasonable computational complexity.\cite{CapMou:07}.} leading to the joint pdf of the measurements and parameters from time $t_1$ to $t_T$ to be
\begin{align}\label{eq:general_model}
	f(\V{z},\B\theta) = 
	\prod_{n = 1}^T f(\B\theta^{(n)}|\B\theta^{(n-1)}) f(\V{z}^{(n)}|\B\theta^{(n)})
\end{align}
where $\B\theta := \B\theta^{(1:T)}$, $\V{z} := \V{z}^{(1:T)}$, and $\B\theta_{0} = \emptyset$ for notational convenience. We consider that (i) the dynamics of different nodes are independent, (ii) the dynamics of positional states are independent of those of measurement parameters, and (iii) the dynamics of different measurement parameters are independent. Therefore, the dynamic model for the parameters in (\ref{eq:general_model}) can be decomposed as
\begin{align}\label{eq:dynamic_model_decom}
	f(\B\theta^{(n)}|\B\theta^{(n-1)}) 
	& = \prod_{k\in\NA} {\bigg[}\,  f\left( \B{\eta}_{k}^{(n)} {\big|} \V{x}_{k}^{(n-1)}, \B{\eta}_{k}^{(n-1)}\right) \\ 
	& \hspace{-15mm}\times
	f\left(\V{x}_{k}^{(n)} {\big|} \V{x}_{k}^{(n-1)}\right) \!\!\! \prod_{j\in\NA\cup\NB\setminus\{k\}}\!\!
f\left(\B{\kappa}_{kj}^{(n)} {\big|} \V{x}_{kj}^{(n-1)}, \B{\kappa}_{kj}^{(n-1)}\right) {\bigg]} \nonumber
\end{align}
where $\V{x}_{kj}^{(n-1)} := \V{x}_{k}^{(n-1)} - \V{x}_{j}^{(n-1)}$. Moreover, since the measurements by different sensors are mutually independent conditioned on the positional state and measurement parameters, the measurement model (likelihood) in (\ref{eq:general_model}) is given by
\begin{align}\label{eq:measurement_model_decom}
	f(\V{z}^{(n)}|\B\theta^{(n)}) 
	& = \prod_{k\in\NA} {\bigg[}\, f\left( \V{z}_{kk}^{(n)}{\big|} \V{x}_{k}^{(n)}, \B{\eta}_{k}^{(n)}\right) \nonumber\\ 
	& \quad \times\!\! \prod_{j\in\NA\cup\NB\setminus\{k\}} f\left(\V{z}_{kj}^{(n)} {\big|} \V{x}_{kj}^{(n)}, \B{\kappa}_{kj}^{(n)}\right) {\bigg]} \,.
\end{align}
Figure \ref{fig:Bayesian} illustrates the graphical model for the above factorization. 

In (\ref{eq:dynamic_model_decom}) and (\ref{eq:measurement_model_decom}), since inter-node measurements depend on the relative positional states, we consider that these measurements and the dynamics of corresponding parameters depend on the {difference} of the positional states between two nodes, i.e., $\V{x}_{kj}^{(n)}$ rather than $\{\V{x}_{k}^{(n)},\V{x}_{j}^{(n)}\}$.

\begin{figure}[t]
	\psfrag{t1}[l][][2.5]{\hspace{-6mm} Time 1}
	\psfrag{t2}[l][][2.5]{\hspace{-6mm} Time 2}
	\psfrag{x3_1}[l][][1.5]{\hspace{-4mm} $\V{x}_3^{(1)}$}
	\psfrag{x1_1}[l][][1.5]{\hspace{-6mm} $\V{x}_1^{(1)}$}
	\psfrag{x2_1}[l][][1.5]{\hspace{-4mm} $\V{x}_2^{(1)}$}
	\psfrag{x3_2}[l][][1.5]{\hspace{-6mm} $\V{x}_3^{(2)}$}
	\psfrag{x1_2}[l][][1.5]{\hspace{-6mm} $\V{x}_1^{(2)}$}
	\psfrag{x2_2}[l][][1.5]{\hspace{-4mm} $\V{x}_2^{(2)}$}

	\psfrag{k12_1}[l][][1.5]{\hspace{-3mm} $\B{\kappa}_{12}^{(1)}$}	
	\psfrag{k12_2}[l][][1.5]{\hspace{-3mm} $\B{\kappa}_{12}^{(2)}$}
	\psfrag{z12_1}[l][][1.5]{\hspace{-3mm} $\V{z}_{12}^{(1)}$}	
	\psfrag{z12_2}[l][][1.5]{\hspace{-3mm} $\V{z}_{12}^{(2)}$}	
	
	\psfrag{k21_1}[l][][1.5]{\hspace{-2mm} $\B{\kappa}_{21}^{(1)}$}	
	\psfrag{k21_2}[l][][1.5]{\hspace{-2mm} $\B{\kappa}_{21}^{(2)}$}
	\psfrag{z21_1}[l][][1.5]{\hspace{-3mm} $\V{z}_{21}^{(1)}$}	
	\psfrag{z21_2}[l][][1.5]{\hspace{-3mm} $\V{z}_{21}^{(2)}$}	

	\psfrag{k13_1}[l][][1.5]{\hspace{-7mm} $\B{\kappa}_{13}^{(1)}$}	
	\psfrag{k13_2}[l][][1.5]{\hspace{-7mm} $\B{\kappa}_{13}^{(2)}$}
	\psfrag{z13_1}[l][][1.5]{\hspace{-4mm} $\V{z}_{13}^{(1)}$}	
	\psfrag{z13_2}[l][][1.5]{\hspace{-4mm} $\V{z}_{13}^{(2)}$}

	\psfrag{k31_1}[l][][1.5]{\hspace{-4mm} $\B{\kappa}_{31}^{(1)}$}	
	\psfrag{k31_2}[l][][1.5]{\hspace{-4mm} $\B{\kappa}_{31}^{(2)}$}
	\psfrag{z31_1}[l][][1.5]{\hspace{-4mm} $\V{z}_{31}^{(1)}$}	
	\psfrag{z31_2}[l][][1.5]{\hspace{-4mm} $\V{z}_{31}^{(2)}$}
	
	\psfrag{k23_1}[l][][1.5]{\hspace{-7mm} $\B{\kappa}_{23}^{(1)}$}	
	\psfrag{k23_2}[l][][1.5]{\hspace{-4mm} $\B{\kappa}_{23}^{(2)}$}
	\psfrag{z23_1}[l][][1.5]{\hspace{-5mm} $\V{z}_{23}^{(1)}$}	
	\psfrag{z23_2}[l][][1.5]{\hspace{-5mm} $\V{z}_{23}^{(2)}$}
	
	\psfrag{k32_1}[l][][1.5]{\hspace{0mm} $\B{\kappa}_{32}^{(1)}$}	
	\psfrag{k32_2}[l][][1.5]{\hspace{0mm} $\B{\kappa}_{32}^{(2)}$}
	\psfrag{z32_1}[l][][1.5]{\hspace{-2mm} $\V{z}_{32}^{(1)}$}	
	\psfrag{z32_2}[l][][1.5]{\hspace{-2mm} $\V{z}_{32}^{(2)}$}
	
	\psfrag{zi1_1}[l][][1.5]{\hspace{-6mm} $\V{z}_{11}^{(1)}$}	
	\psfrag{zi1_2}[l][][1.5]{\hspace{-6mm} $\V{z}_{11}^{(2)}$}
	\psfrag{zi2_1}[l][][1.5]{\hspace{-5mm} $\V{z}_{22}^{(1)}$}	
	\psfrag{zi2_2}[l][][1.5]{\hspace{-5mm} $\V{z}_{22}^{(2)}$}
	\psfrag{zi3_1}[l][][1.5]{\hspace{-5mm} $\V{z}_{33}^{(1)}$}	
	\psfrag{zi3_2}[l][][1.5]{\hspace{-5mm} $\V{z}_{33}^{(2)}$}
	\psfrag{n1_1}[l][][1.5]{\hspace{-3.5mm} $\B\eta_{1}^{(1)}$}
	\psfrag{n1_2}[l][][1.5]{\hspace{-3.5mm} $\B\eta_{1}^{(2)}$}
	\psfrag{n2_1}[l][][1.5]{\hspace{-3.5mm} $\B\eta_{2}^{(1)}$}
	\psfrag{n2_2}[l][][1.5]{\hspace{-3.5mm} $\B\eta_{2}^{(2)}$}
	\psfrag{n3_1}[l][][1.5]{\hspace{-5mm} $\B\eta_{3}^{(1)}$}	
	\psfrag{n3_2}[l][][1.5]{\hspace{-5mm} $\B\eta_{3}^{(2)}$}
	
	\centering
	{\includegraphics[angle=0,width=0.95\linewidth,draft=false]{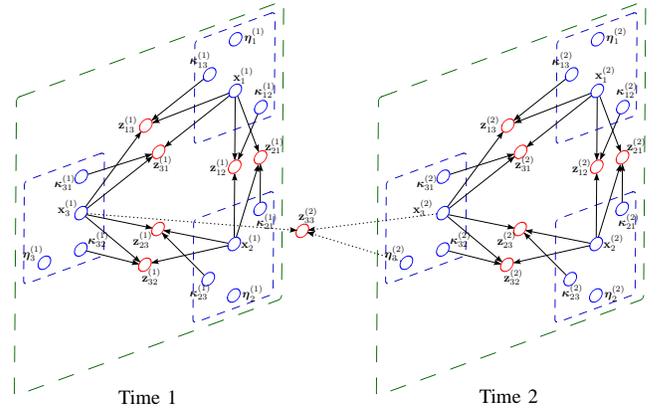}}

    \caption{Graphical model describing the interrelationships among the parameters and measurements present in the deterministic model for cooperative network navigation. The temporal cooperation is illustrated only for agent 3 with $n_0=1$ and $n_0'=0$. \label{fig:deterministic}}
\end{figure}

\subsubsection{Deterministic model}\label{sec:Deterministic_model}

The agents' positional states and the measurement parameters can also be described as deterministic unknowns. In this case, there is no prior dynamic knowledge, and the measurements depend on the agents' positions and orientations in a set of consecutive time instants. Hence, the positional state $\V{x}_{k}^{(n)}$ includes only the position and orientation of agent $k$, and the likelihood for the deterministic model can be written as\footnote{Throughout the paper, the notation ${(n-n_0:n+n_0')}$ is shorthand for ${(\max\{n-n_0,1\}:\min\{n+n_0',T\})}$.}
\begin{align}\label{eq:nonbayesian_model}
	f(\V{z}|\B\theta)
	& = \prod_{n = 1}^{T}  \prod_{k\in\NA} {\bigg[} f\left(\V{z}_{kk}^{(n)}{\big|} \V{x}_{k}^{(n-n_0:n+n_0')}, \B{\eta}_{k}^{(n)}\right) 
	 \\ & \qquad \quad  \times \!\!
	 \prod_{j\in\NA\cup\NB\setminus\{k\}} f\left( \V{z}_{kj}^{(n)} {\big|} \V{x}_{kj}^{(n-n_0:n+n_0')}, \B{\kappa}_{kj}^{(n)}\right) {\bigg]} \nonumber
\end{align}
where the measurements at time $t_n$ depend only on the agents' positions and orientations from $t_{n-n_0}$ to $t_{n+n_0'}$ as well as the measurement parameters at $t_n$, and the same independence condition of the measurements is used as in the Bayesian model (see Fig.~\ref{fig:deterministic} for the graphical model). 

Note that the difference between the Bayesian and deterministic models is that the latter (i) considers all parameters to be deterministic unknowns, i.e., assumes no prior knowledge about dynamics, and (ii) includes only the position and orientation in the positional state and the measurements depend on such positional states in consecutive time instants. For example, when the intra-node measurement is velocity, the Bayesian model directly employs the velocity in the positional state, while the deterministic model can formulate the measurement as $f\left(\V{z}_{kk}^{(n)}{\big|} \V{p}_{k}^{(n-1:n)}, \B{\eta}_{k}^{(n)}\right)$ determined by
\begin{align*}
	\V{z}_{kk}^{(n)} = \frac{\V{p}_{k}^{(n)}-\V{p}_{k}^{(n-1)}}{\Delta_t} + \V{w}_{k}^{(n)}
\end{align*}
where $\Delta_t$ is the sampling time interval, and $\V{w}_{k}^{(n)}$ is the residual error. 

\begin{figure*}
	[!b] \vspace*{4pt} 
	\hrulefill
	\normalsize%
	\setcounter{MYtempeqncnt}{\value{equation}} 
	\setcounter{equation}{11} 
	\begin{align}\label{eq:thm_s_n} 
		\V{S}^{(n,m)} & = \Matrix{cccccc}{\sum\limits_{j\in\NA\cup\NB\setminus\{1\}}
		\V{S}_{1,j}^{(n,m)} & -\V{S}_{1,2}^{(n,m)} & \cdots &  -\V{S}_{1,\Na}^{(n,m)}\\ 
		-\V{S}_{1,2}^{(n,m)\,\dagger} & \sum\limits_{j\in\NA\cup\NB\setminus\{2\}}
		\V{S}_{2,j}^{(n,m)} & & -\V{S}_{2,\Na}^{(n,m)} \\
		\vdots & & \ddots \\
		-\V{S}_{1,\Na}^{(n,m)\,\dagger}  & -\V{S}_{2,\Na}^{(n,m)\, \dagger}  & &  \sum\limits_{j\in\NA\cup\NB\setminus\{\Na\}} \V{S}_{\Na,j}^{(n,m)}} 
	\end{align}
		\setcounter{equation}{\value{MYtempeqncnt}} \vspace*{-6pt} 
\end{figure*}

\subsection{Equivalent Fisher Information}

The mean squared error (MSE) matrix of an estimate $\hat{\V{x}}_k$ for the $k$th agent's positional state is bounded below as\footnote{With a slight abuse of notation, $\E_{\V{z},\B\theta}\{\cdot\}$ in \eqref{eq:def_SPEB2} will be used for deterministic, random, and hybrid cases, with the understanding that the expectation operation is not performed over the deterministic components of $\B\theta$ \cite{Poo:B94, ReuMes:97}. 
} 
    \begin{align}\label{eq:def_SPEB2}
        \E_{\V{z}, \B\theta} \left\{ ( \hat{\V{x}}_k^{(n)} - \V{x}_k^{(n)} )
        (\hat{\V{x}}_k^{(n)} - \V{x}_k^{(n)} )^\dagger \right\}
        \succeq  \left[ \J_{\B\theta}^{-1} \right]_{\V{x}_k^{(n)}} 
    \end{align}
where $\J_{\B\theta}$ is the Fisher information matrix (FIM) for $\B\theta$ and $\left[ \J_{\B\theta}^{-1} \right]_{\V{x}_k^{(n)}}$ denotes the square submatrix on the diagonal of $\J_{\B\theta}^{-1}$ corresponding to $\V{x}_k^{(n)}$ \cite{SheWin:J10a}. 
Since only a small submatrix of $\J_{\B\theta}^{-1}$ corresponding to the positional states is of interest, we apply the EFI analysis to reduce the dimension of the FIM while retaining all the information for the parameters of interest \cite{SheWin:J10a}.

\begin{defi}[EFIM]\label{def:EFIM}
Given a parameter vector $\B\theta = [\, \B\theta_1^\dagger \;\;
\B\theta_2^\dagger \,]^\dagger$ and the FIM $\J_{\B\theta}$ of the form
    \begin{align*}
        \J_{\B\theta} = \Matrix{cc}
                {   \V{A}         &   \V{B}   \\
                    \V{B}^\dagger &   \V{C}   },
    \end{align*}
where $\bm\theta \in \mathbb{R}^N$, $ \B\theta_1 \in \mathbb{R}^n$, $\V{A} \in \mathbb{R}^{n \times n}$, $\V{B} \in \mathbb{R}^{n\times (N-n)}$, and $\V{C} \in \mathbb{R}^{(N-n) \times (N-n)}$ with $1\leq n<N$, the EFIM for $\bm\theta_1$ is given by
    \begin{align}\label{eq:Sing_EFIM_FIM}
        \JE{\B\theta_1} := \V{A} - \V{B} \V{C}^{-1} \V{B}^\dagger
        \, .
    \end{align}
\end{defi}

We call matrix $\V{B}$ the \emph{cross-information} between $\B\theta_1$ and $\B\theta_2$ in $\J_{\B\theta}$. The right-hand side of (\ref{eq:Sing_EFIM_FIM}) is known as the Schur's complement of matrix $\V{A}$ \cite{HorJoh:B85}. 
The EFIM retains all the necessary information to derive the information inequality \cite{Tre:68} for the parameter $\B\theta_1$, in the sense that $[\,\J_{\B\theta} ^{-1}]_{\B\theta_1} = \left[\, \JE{\B\theta_1} \, \right]^{-1}$. 

\section{EFIM for Network Navigation}\label{sec:EFIM_Navigation}

In this section, we first derive the EFIMs for the agents' positional states based on both Bayesian and deterministic navigation models. We then discuss the implications of the EFIMs as well as the challenges in realizing network navigation. For notational convenience, we denote $\V{x} := \V{x}_{1:\Na}^{(1:T)}$, $\B{\eta} := \B{\eta}_{1:\Na}^{(1:T)}$, $\B{\kappa} := \B{\kappa}_{1:\Na}^{(1:T)}$, and functionals
\begin{align*}
	\B\Phi_{\B\alpha,\B\beta}(f) & := \E_{\V{z},\B\theta} \left\{ -\frac{\partial^2 \ln f}{\partial \B\alpha \, \partial \B\beta^\dagger} \right\} \\
	\B\Psi_{\B\alpha,\B\beta}^{\B\gamma} (f) & := \B\Phi_{\B\alpha,\B\beta}(f)  
	- \B\Phi_{\B\alpha,\B\gamma}(f) \left[\B\Phi_{\B\gamma,\B\gamma}(f)\right]^{-1} \B\Phi_{\B\gamma,\B\beta}(f) \,.
\end{align*}

\subsection{EFIM for Bayesian Model}

We first consider the Bayesian formulation of cooperative network navigation as introduced in Section \ref{sec:Bayesian_model} and derive the EFIM for the agents' positional states based on (\ref{eq:general_model}). In the following theorem, we show that the EFIM can be decomposed into the sum of information corresponding to the mobility model, temporal cooperation, and spatial cooperation. This result implies that each source of cooperation or prior knowledge contributes to navigation information in an additive way. 

\begin{thm}\label{thm:EFIM_complete}
For the Bayesian model of network navigation, the EFIM for the agents' positional states $\V{x}$ from time $t_1$ to $t_T$ is given by
\begin{align}\label{eq:EFIM_complete}
	\J_\text{e}(\V{x}) = \J_\text{e}^\text{m}(\V{x}) + \J_\text{e}^\text{t}(\V{x}) + \J_\text{e}^\text{s}(\V{x}) 
\end{align}
where $\J_\text{e}^\text{m}(\V{x})$ is the EFIM corresponding to the mobility model, given by a diagonally-striped matrix
\begin{align}\label{eq:thm_Jp}
	\J_\text{e}^\text{m}(\V{x}) 
	= \Matrix{ccccc}{\V{P}^{(1,1)} & \V{P}^{(1,2)}\\ 
	\V{P}^{(1,2)\,\dagger} &\V{P}^{(2,2)} & \V{P}^{(2,3)}\\ 
	& \V{P}^{(2,3)\,\dagger} & \ddots \\ 
	&&&\V{P}^{(T,T)}}
\end{align}
in which $\V{P}^{(n,m)} = \text{diag}{\big\{} \V{P}^{(n,m)}_1, \V{P}^{(n,m)}_2, \ldots, \V{P}^{(n,m)}_\Na{\big\}}$ with $\V{P}^{(n,m)}_k$ given by (\ref{eq:p_n_m}); $\J_\text{e}^\text{t}(\V{x})$ is the EFIM  corresponding to temporal cooperation, given by
\begin{align}\label{eq:thm_Jt}
	\J_\text{e}^\text{t}(\V{x}) = \Matrix{cccccc}{ \V{K}^{(1,1)} & \V{K}^{(1,2)} & \cdots &  \V{K}^{(1,T)}\\ 
	\V{K}^{(1,2)\,\dagger} & \V{K}^{(2,2)} & & \V{K}^{(2,T)}\\
	\vdots & & \ddots \\
	\V{K}^{(1,T)\,\dagger} & \V{K}^{(2,T)\,\dagger} & &\V{K}^{(T,T)}}
\end{align}
in which $\V{K}^{(n,m)} = \text{diag}{\big\{} \V{K}^{(n,m)}_1, \V{K}^{(n,m)}_2,\ldots, \V{K}^{(n,m)}_\Na{\big\}}$ with $\V{K}^{(n,m)}_k$ given by (\ref{eq:K_n_m}); and $\J_\text{e}^\text{s}(\V{x})$ is the EFIM corresponding to spatial cooperation, given by
\begin{align}\label{eq:thm_Js}
	\J_\text{e}^\text{s}(\V{x}) = \Matrix{cccccc}{ \V{S}^{(1,1)} & \V{S}^{(1,2)} & \cdots &  \V{S}^{(1,T)}\\ 
	\V{S}^{(1,2)\,\dagger} & \V{S}^{(2,2)} & & \V{S}^{(2,T)}\\
	\vdots & & \ddots \\
	\V{S}^{(1,T)\,\dagger} & \V{S}^{(2,T)\,\dagger} & &\V{S}^{(T,T)}}
\end{align}
in which $\V{S}^{(n,m)}$ is given by (\ref{eq:thm_s_n}) shown at the bottom of the page, with $\V{S}^{(n,m)}_{kj}$ given by (\ref{eq:Gkj_n_m}).
 \addtocounter{equation}{1}
\end{thm}

\begin{IEEEproof}
See Appendix \ref{apd:proof_EFIM_complete} for the outline of the proof.\footnote{Note that detailed proofs are omitted throughout the paper due to the space constraints.}
\end{IEEEproof}

\begin{remrk}
Theorem \ref{thm:EFIM_complete} shows that the EFIM for the positional states can be decomposed into three parts, i.e., the information corresponding to the mobility model, temporal cooperation, and spatial cooperation. Each part has a specific structure explained in the following:
\begin{itemize}
	\item Mobility model: $\J_\text{e}^\text{m}(\V{x})$ characterizes the navigation information, i.e., information about the positional states, corresponding to the mobility model. Every  matrix $\V{P}^{(n,n)}$ on the diagonal of $\J_\text{e}^\text{m}(\V{x})$ is block-diagonal with each block of size $D\times D$ corresponding to $\V{x}^{(n)}_k$'s, since the mobilities of different agents are independent. Moreover, since the mobility model characterizes the positional states in two consecutive time instants, only the matrices $\V{P}^{(n,n+1)}$ in off-diagonals of $\J_\text{e}^\text{m}(\V{x})$ are non-zero. Furthermore, the matrices $\V{P}^{(n,n+1)}$ are block-diagonal with each block of size $D\times D$ corresponding to the statistical distribution of mobility $f(\V{x}^{(n+1)}_k|\V{x}^{(n)}_k)$, again since the mobilities of different agents at different time instants are independent.
	
	\item Temporal cooperation: $\J_\text{e}^\text{t}(\V{x})$ characterizes the navigation information corresponding to the cooperation via intra-node measurements. Every matrix $\V{K}^{(n,n)}$ on the diagonal of $\J_\text{e}^\text{t}(\V{x})$ is block-diagonal with blocks of size $D\times D$ corresponding to $\V{x}^{(n)}_k$'s, since the intra-node measurements of different agents are independent. Furthermore, the off-diagonal matrices $\V{K}^{(n,m)}$ are block-diagonal with each block $\V{K}^{(n,m)}_k$ of size $D\times D$ corresponding to the cross-information of the positional states $\V{x}^{(n)}_k$ and $\V{x}^{(m)}_k$ in $\J_\text{e}^\text{t}(\V{x})$. This cross-information arises from the elimination of the parameters $\B\eta$. The absence of cross-information between the positional states of different agents, i.e., the zero cross-information  for $\V{x}^{(n)}_k$ and $\V{x}^{(m)}_j$ where $(k\neq j)$, is due to the independence of all the intra-node measurements among the agents.

	
	\item Spatial cooperation: $\J_\text{e}^\text{s}(\V{x})$ characterizes the navigation information corresponding to the cooperation among the agents. Every  matrix $\V{S}^{(n,n)}$ on the diagonal of $\J_\text{e}^\text{s}(\V{x})$ has the specific structure shown in (\ref{eq:thm_s_n}): (i) the submatrix $\V{S}^{(n,n)}_{kj}$ characterizes the navigation information obtained from the inter-node measurement between node $k$ and $j$, (ii) the $k$th block on the diagonal is the summation of such information between agent $k$ and all other nodes, and (iii) the off-diagonal blocks are the negative of such information between each pair of agents due to the uncertainty of the agents's positional states in cooperation \cite{SheWymWin:J10}. The off-diagonal matrices $\V{S}^{(n,m)}$ in $\J_\text{e}^\text{s}(\V{x})$ have the same structure as $\V{S}^{(n,n)}$, and these matrices arise due to the elimination of the parameters $\B\kappa$.
\end{itemize}
\end{remrk}

Moreover, one can show from Theorem \ref{thm:EFIM_complete} that when an agent has infinite Fisher information about its positional states, its role in the navigation network is the same as an anchor. Therefore, there is no fundamental difference between anchors and agents, and our framework unifies these nodes in the network from the Bayesian point of view.

Theorem \ref{thm:EFIM_complete} presents the EFIM for network navigation based on the general Bayesian model. When the measurement parameters have unknown or simple dynamics, they can be modeled as mutually independent of the positional states and for different time instants, in which case $f{\big(} \B{\eta}_{k}^{(n)} {\big|} \V{x}_{k}^{(n-1)}, \B{\eta}_{k}^{(n-1)} {\big)} = f{\big(}\B{\eta}_{k}^{(n)} {\big)}$ and $f{\big(}\B{\kappa}_{kj}^{(n)} {\big|} \V{x}_{kj}^{(n-1)}, \B{\kappa}_{kj}^{(n-1)}{\big)}= f {\big(} \B{\kappa}_{kj}^{(n)} {\big)}$ in (\ref{eq:dynamic_model_decom}). We derive the EFIM for this special case in the following corollary. 

\begin{cor}\label{thm:EFIM_independent}
When the measurement parameters are mutually independent of the positional states and mutually independent for different time instants, the EFIM for the agents' positional states ${\V{x}}$ from time $t_1$ to $t_T$ is given by
\begin{align}\label{eq:EFIM_cor_1}
	\J_\text{e}(\V{x}) = \J_\text{e}^\text{m}(\V{x}) + \J_\text{e}^\text{t}(\V{x}) + \J_\text{e}^\text{s}(\V{x}) 
\end{align}
where $\J_\text{e}^\text{m}(\V{x})$ is given by (\ref{eq:thm_Jp}); $\J_\text{e}^\text{t}(\V{x}) = \text{diag}{\big\{}\V{K}^{(1,1)}, \V{K}^{(2,2)},\ldots, \V{K}^{(T,T)}{\big\}}$ is block-diagonal, 
with $\V{K}^{(n,n)}=\text{diag}{\big\{}\V{K}^{(n,n)}_1, \V{K}^{(n,n)}_2,\ldots,\V{K}^{(n,n)}_\Na{\big\}}$; and $\J_\text{e}^\text{s}(\V{x}) = \text{diag} {\big\{} \V{S}^{(1,1)}, \V{S}^{(2,2)},\ldots, \V{S}^{(T,T)}{\big\}}$ is also block-diagonal, 
in which $\V{S}^{(n,n)}$ has the same structure as given in (\ref{eq:thm_s_n}).
In these equations, the matrix $\V{P}_k^{(n,n)}$ is given by (\ref{eq:p_n_m}), and
\begin{align}\label{eq:kk_n_1}
	\V{K}_k^{(n,n)} & =  \B\Psi_{\V{x}_{k}^{(n)}, \V{x}_{k}^{(n)}}^{\B\eta_{k}^{(n)}}\left(f(\B\eta_{k}^{(n)}) \cdot f(\V{z}_{kk}^{(n)}|\V{x}_{k}^{(n)},\B\eta_{k}^{(n)})\right)  \\
\label{eq:skj_n_1}
	\V{S}_{kj}^{(n,n)} & = \B\Psi_{\V{x}_{k}^{(n)}, \V{x}_{k}^{(n)}}^{\B\kappa_{kj}^{(n)}} \left( f(\B\kappa_{kj}^{(n)}) \cdot f(\V{z}_{kj}^{(n)}|\V{x}_{kj}^{(n)},\B\kappa_{kj}^{(n)}) \right) \nonumber \\ & \quad + \B\Psi_{\V{x}_{k}^{(n)}, \V{x}_{k}^{(n)}}^{\B\kappa_{jk}^{(n)}} \left(f(\B\kappa_{jk}^{(n)}) \cdot f(\V{z}_{jk}^{(n)}|\V{x}_{jk}^{(n)},\B\kappa_{jk}^{(n)}) \right)\,.
\end{align}
\end{cor}

\begin{IEEEproof}
(Outline) Due to the independence condition, off-diagonal submatrices or cross-information in $\J_{\B\theta}$ are zeros, e.g., $\B\Phi_{\B\eta_{k}^{(n)}, \B\eta_{k}^{(n+1)}}{\big(}  f(\B\eta_{k}^{(n+1)}|\V{x}_{k}^{(n)}, \B\eta_{k}^{(n)}){\big)}= \V{0}$. Based on this fact, (\ref{eq:kk_n_1}) and (\ref{eq:skj_n_1}) can be obtained after some algebra. 
\end{IEEEproof}

\begin{remrk}
Under the independence condition of the measurement parameters, Corollary \ref{thm:EFIM_independent} shows that both the EFIMs corresponding to the temporal and spatial cooperation are block-diagonal. In other words, since the correlation between the measurement parameters in time no longer exists, the inter- and intra-node measurements do not induce any cross-information in the  navigation information for different time instants.
\end{remrk}

\subsection{EFIM for Deterministic Model}\label{sec:EFIM_det}

We now consider all the parameters to be deterministic unknowns as introduced in Section \ref{sec:Deterministic_model}. In the following theorem, we derive the corresponding EFIM for the agents' positional states, i.e., positions and orientations in this case, based on the model in (\ref{eq:nonbayesian_model}). The decomposition of the EFIM is analogous to that in the previous section, but with the absence of the term corresponding to the mobility model.

\begin{thm}\label{thm:EFIM_deterministic}
For the deterministic model of network navigation, the EFIM for the agents' positional states $\V{x}$ from time $t_1$ to $t_T$ is given by 
\begin{align}\label{eq:EFIM_det}
	\J_\text{e}(\V{x}) = \J_\text{e}^\text{t}(\V{x}) + \J_\text{e}^\text{s}(\V{x}) 
\end{align}
where $\J_\text{e}^\text{t}(\V{x})$ is the EFIM from temporal cooperation, structured as (\ref{eq:thm_Jt}),
and $\J_\text{e}^\text{s}(\V{x})$ is the EFIM from spatial cooperation, structured as (\ref{eq:thm_Js}),
in which the block matrices $\V{K}^{(n,m)}_k$ and $\V{S}_{kj}^{(n,m)}$ are given by 
	\begin{align}\label{eq:det_K_n_m}
		\V{K}_k^{(n,m)} 
		& =  \begin{cases}
		\sum\limits_{l=m-n_0'}^{n+n_0} \B\Psi_{\V{x}_{k}^{(n)}, \V{x}_{k}^{(m)}}^{\B\eta_{k}^{(l)}} \left( f(\V{z}_{kk}^{(l)}|\V{x}_{k}^{(l-n_0:l+n_0')},\B\eta_{k}^{(l)}) \right)
		\,,  \\
		& \hspace{-36mm} n \leq m \leq n+ n_0+n_0' \\
		\V{0} \,,  &  \hspace{-36mm}  m > n+n_0+n_0'
	\end{cases}
	\end{align}
	\begin{align}\label{eq:det_Gkj_n_m}
		\V{S}_{kj}^{(n,m)} 
		& = \begin{cases}
			\sum\limits_{l=m-n_0'}^{n+n_0} {\bigg[}  \B\Psi_{\V{x}_{k}^{(n)}, \V{x}_{k}^{(m)}}^{\B\kappa_{kj}^{(l)}} \left( f(\V{z}_{kj}^{(l)}|\V{x}_{kj}^{(l-n_0:l+n_0')},\B\kappa_{kj}^{(l)}) \right) \\ \hspace{8mm}  + \B\Psi_{\V{x}_{k}^{(n)}, \V{x}_{k}^{(m)}}^{\B\kappa_{jk}^{(l)}} \left( f(\V{z}_{jk}^{(l)}|\V{x}_{jk}^{(l-n_0:l+n_0')},\B\kappa_{jk}^{(l)})\right)
			{\bigg]} \,, \\
			& \hspace{-36mm} n \leq m \leq n+ n_0+n_0' \\
			\V{0}  \,,& \hspace{-36mm}  m > n+n_0+n_0' \,.
		\end{cases}
	\end{align}
\end{thm}


\begin{figure*}
	[!b] \vspace*{4pt} 
	\hrulefill
	\normalsize%
	\setcounter{MYtempeqncnt}{\value{equation}} 
	\setcounter{equation}{18} 
	\begin{align}\label{eq:example_1}
	& \J_\text{e}({\V{p}}) =  
	\Matrix{cccccc}{ {\V{S}}^{(1)}+{\V{K}}^{(2)}& -{\V{K}}^{(2)} \\ -{\V{K}}^{(2)}
	& {\V{S}}^{(2)}+{\V{K}}^{(2)}+{\V{K}}^{(3)} & -{\V{K}}^{(3)} \\ & -{\V{K}}^{(3)} & {\V{S}}^{(3)} + {\V{K}}^{(3)} + {\V{K}}^{(4)}\\ &&\ddots&\ddots \\ &&& {\V{S}}^{(T-1)}+{\V{K}}^{(T-1)}+{\V{K}}^{(T)} & -{\V{K}}^{(T)} \\ &&& -{\V{K}}^{(T)} & {\V{S}}^{(T)}+{\V{K}}^{(T)}}
	\end{align}
	\setcounter{equation}{\value{MYtempeqncnt}} \vspace*{-6pt} 
\end{figure*}

\begin{remrk}
The proof of the theorem follows from a similar derivation of Theorem \ref{thm:EFIM_complete}. Compared to the EFIM for the Bayesian case in (\ref{eq:EFIM_complete}) of Theorem \ref{thm:EFIM_complete}, the EFIM for the deterministic case in (\ref{eq:EFIM_det}) does not contain the components corresponding to the dynamic models for the mobility and measurement parameters. Theorem \ref{thm:EFIM_deterministic} shows that the EFIM in (\ref{eq:EFIM_det}) can be decomposed into two parts, i.e., the information from temporal and spatial cooperation, where $\J_\text{e}^\text{t}(\V{x})$ and $\J_\text{e}^\text{s}(\V{x})$ as well as their submatrices have the same decomposition as their counterparts in Theorem \ref{thm:EFIM_complete}. Hence, the remarks drawn for the Bayesian case on temporal and spatial cooperation can be extended to this case. 

Specifically for this case, since the measurements made at time $t_n$ are only related to the positional states from $t_{n-n_0}$ to $t_{n+n_0'}$, the cross-information $\V{K}_k^{(n,m)}$ and $\V{S}_{kj}^{(n,m)}$ between $\V{x}_k^{(n)}$ and $\V{x}_k^{(m)}$ depend only on the measurements from $t_{m-n_0'}$ to $t_{n+n_0}$. In other words, this cross-information is zero for pairs of positional states whose time span is larger than $n_0+n_0'$, i.e., $\V{K}_k^{(n,m)}=\V{0}$ for $m>n+n_0+n_0'$.
\end{remrk}

\subsection{Discussion}

The above results show that joint temporal and spatial cooperation can significantly improve the navigation performance, since each type of cooperation corresponds to a positive semi-definite matrix in the sum for the total EFIM. However, realizing this benefit in practical implementation incurs associated costs such as additional communication and complex information fusion: (i) the communication among nodes required for cooperation can jeopardize the benefits of cooperation if such communication is not properly designed \cite{WinConMazSheGifDarChi:J11,KhaKarMou:09}, and (ii) the non-diagonal structure of the above EFIMs implies strong correlation in agents' position estimates and hence hinders the development of distributed algorithms for cooperative information fusion that are favored in medium- and large-scale networks \cite{WinConMazSheGifDarChi:J11, KhaMou:08, MazSheWin:L11}. Hence, for realistic network design and operation, it is crucial to develop efficient communication strategies including message representation and scheduling techniques, tailored  specifically to network navigation, as well as efficient information fusion techniques that can combine information from various cooperation sources in a distributed manner. 

To handle the trade-off between performance and complexity, it is essential to grasp the essence of the network navigation process. To this end and to gain insights for network design and operation, we will further explore the implications of the navigation information and develop a geometrical interpretation of information evolution.

\section{Geometrical Interpretation of Information Evolution} \label{sec:geometrical_interpretation}

In this section, we investigate the EFIM for the agents' positions in 2-D navigation networks, and develop a geometrical interpretation of navigation information to illustrate the information evolution in navigation.


\subsection{EFIM for 2-D Navigation Network}

We focus on the positions of the agents in a 2-D navigation network and assume their orientations are known for simplicity, i.e., the positional state only includes the position and its derivatives. Each agent obtains intra-node measurements for its velocity and inter-node measurements for the distances to its neighboring nodes.\footnote{For example, the agent's velocity can be measured by inertial sensors or Doppler radar, and the ranges can be measured from RF signals transmitted between nodes. 
} 

The intra-node measurement is modeled as $f({\V{z}}_{kk}^{(n)}|\V{x}_k^{(n)},\B\eta_k^{(n)})$ for the Bayesian case and as $f({\V{z}}_{kk}^{(n)}|\V{p}_k^{(n)}-\V{p}_k^{(n-1)},\B\eta_k^{(n)})$ for the deterministic case, whereas the inter-node measurement is modeled as $f({\V{z}}_{kj}^{(n)}| \|\V{p}_{k}^{(n)}-\V{p}_{j}^{(n)}\|, \B\kappa_{kj}^{(n)})$. It can be shown that the EFIM for the agents' positions
for the deterministic case has the same structure as that for the Bayesian case if the mobility model includes only the agents' positions. Hence, without loss of generality, we investigate the EFIM for the deterministic case in this section. 

Let the angle $\phi^{(n)}_{kk}$ describe the direction from $\V{p}_k^{(n-1)}$ to $\V{p}_k^{(n)}$ and $\phi^{(n)}_{kj}$ describe the direction from $\V{p}_{k}^{(n)}$ to $\V{p}_{j}^{(n)}$; $d_{kk}^{(n)} := \|\V{p}_{k}^{(n)} - \V{p}_{k}^{(n-1)}\|$ and $d_{kj}^{(n)} := \|\V{p}_{k}^{(n)} - \V{p}_{j}^{(n)} \|$; and $\R(\phi) :=  \V{u}_{\phi} \V{u}_{\phi}^\dagger$ and $\R(\phi,\phi+\pi/2) :=  {\big[}\V{u}_{\phi} \, {\V{u}}_{\phi}^{\perp\,\dagger} + {\V{u}}_{\phi}^\perp \, \V{u}_{\phi}^\dagger {\big]}/2$. The next theorem obtains the simple structure of the EFIM, showing that there is only cross-information between consecutive time instants and the information from each measurement is characterized by a $2\times 2$ matrix.

\begin{figure*}
	[!b] \vspace*{4pt} 
	\hrulefill
	\normalsize%
	\setcounter{MYtempeqncnt}{\value{equation}} 
	\setcounter{equation}{19} 
	\begin{align}\label{eq:EFIM_carry-over}
		&  \J_\text{e}({\V{p}}^{(n:T)}) = 
		\Matrix{cccccc}{ {\V{S}}^{(n)}+\widetilde{\V{K}}^{(n)}+{\V{K}}^{(n+1)} & -{\V{K}}^{(n+1)} \\ 
		-{\V{K}}^{(n+1)} & {\V{S}}^{(n+1)} + {\V{K}}^{(n+1)} + {\V{K}}^{(n+2)}\\ 
		& \ddots & \ddots \\ && {\V{S}}^{(T-1)}+{\V{K}}^{(T-1)}+{\V{K}}^{(T)} & -{\V{K}}^{(T)} \\ 
		&& -{\V{K}}^{(T)} & {\V{S}}^{(T)}+{\V{K}}^{(T)}}
	\end{align}
	\setcounter{equation}{\value{MYtempeqncnt}} \vspace*{-6pt} 
\end{figure*}

\begin{thm}\label{thm:example_1}
For the deterministic model of 2-D network navigation with measurements of the agent velocity and inter-node distance, the EFIM for the agents' positions $\V{p}$ from time $t_1$ to $t_T$ is given by (\ref{eq:example_1}) shown at the bottom of the page, \addtocounter{equation}{1}
where the EFIM from temporal cooperation is ${\V{K}}^{(n)} = \text{diag} {\big\{}\V{K}_{1}^{(n)},\V{K}_{2}^{(n)},\ldots, \V{K}_{\Na}^{(n)}{\big\}}$ and the EFIM from spatial cooperation is ${\V{S}}^{(n)}$ structured as (\ref{eq:thm_s_n}).\footnote{The notations ${\V{S}}^{(n)}$, $\V{S}_{kj}^{(n)}$, $\V{K}^{(n)}$, and $\V{K}_{k}^{(n)}$ are shorthands for ${\V{S}}^{(n,n)}$, $\V{S}_{kj}^{(n,n)}$, $\V{K}^{(n,n)}$, and $\V{K}_{k}^{(n,n)}$, respectively.}
In the above expressions, $\V{K}_{k}^{(n)}, \V{S}^{(n)}_{kj} \in \mathbb{S}_{++}^2$ are given by
\begin{align*}
	\V{K}_{k}^{(n)} & = \lambda^{(n)}_{kk} \cdot \R(\phi^{(n)}_{kk}) + \nu^{(n)}_{kk} \cdot \R(\phi^{(n)}_{kk}+\pi/2) \\ & \quad + \xi^{(n)}_{kk} \cdot \R(\phi^{(n)}_{kk},\phi^{(n)}_{kk}+\pi/2)  \\[1ex]
	\V{S}^{(n)}_{kj} & = \lambda^{(n)}_{kj} \cdot \R(\phi_{kj}^{(n)})
\end{align*}
where $\lambda^{(n)}_{kk} = [\breve{\V{K}}_{k}^{(n)}]_{1,1}$, $\xi^{(n)}_{kk} = [\breve{\V{K}}_{k}^{(n)}]_{1,2}/d^{(n)}_{kk}$, $\nu^{(n)}_{kk} = [\breve{\V{K}}_{k}^{(n)}]_{2,2}/d^{(n)\;2}_{kk}$, and $\lambda^{(n)}_{kj} = \breve{S}^{(n)}_{kj}$, with
\begin{align*}
	\breve{\V{K}}_{k}^{(n)} & = \B\Psi_{[\,d^{(n)}_{kk}\; \phi^{(n)}_{kk}\,], [ d^{(n)}_{kk}\; \phi^{(n)}_{kk} ]}^{\B\eta_k^{(n)}} \left( f({\V{z}}_{kk}^{(n)}| d^{(n)}_{kk}, \phi^{(n)}_{kk} ,\B\eta_k^{(n)})\right) \\
	\breve{S}^{(n)}_{kj} & = \B\Psi_{d_{kj}^{(n)},d_{kj}^{(n)}}^{\B\kappa_{kj}^{(n)}}\left( f({\V{z}}_{kj}^{(n)}|d_{kj}^{(n)} ,\B\kappa_{kj}^{(n)})\right) \,.
\end{align*}
In particular, $\xi^{(n)}_{kk}=0$ when the amplitude and direction measurements of the velocity are independent.
\end{thm}


\begin{figure}[t]
	\vspace*{8mm}
	\centering
	\subfigure[Temporal Cooperation]{
	\psfrag{time1}[l][][1.5]{\hspace{-6mm} Time 1}
	\psfrag{time2}[l][][1.5]{\hspace{-6mm} Time 2}
	\psfrag{J11}[l][][1.5]{\hspace{-5mm} $\V{S}_1^{(1)}$}
	\psfrag{phi12}[l][][1.5]{\hspace{-4mm} $\phi_{1,2}$}
	
	\psfrag{J21}[l][][1.5]{\hspace{-5mm} $\V{S}_{1}^{(2)}$}
	\psfrag{v}[l][][1.5]{\hspace{-1.5mm} $\V{D}_{2}$}
	\psfrag{v2}[l][][1.5]{\hspace{-2mm} $\V{C}_{2}$}
	\psfrag{J22}[l][][1.5]{\hspace{-3mm} $\V{J}_\text{e}(\V{p}_1^{(2)})$}
	 \includegraphics[angle=0,width=0.46\linewidth,draft=false]{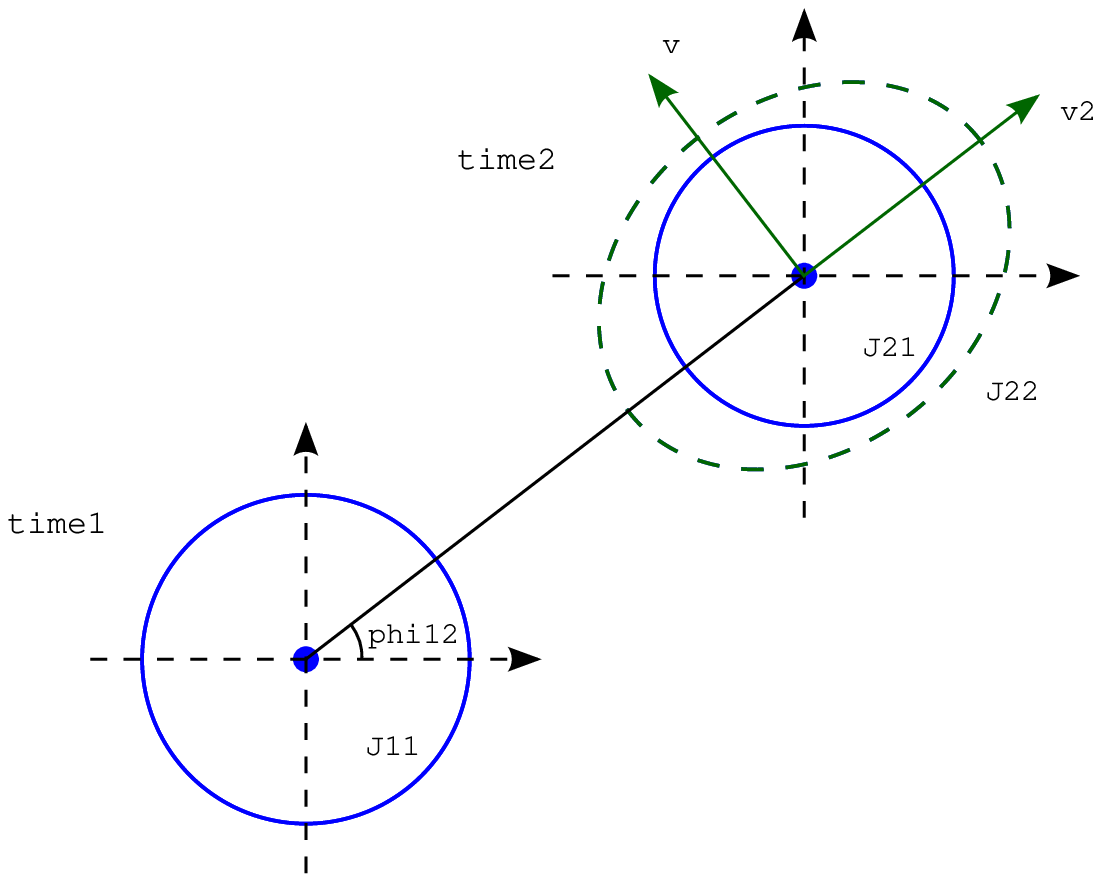} \label{fig:temporal_coop}}
	\hfill
	\subfigure[Spatial Cooperation]{
	\psfrag{Agent 1}[l][][1.5]{\hspace{-10mm} Agent 1}
	\psfrag{Agent 2}[l][][1.5]{\hspace{-10mm} Agent 2}
	
	\psfrag{phi12}[l][][1.5]{\hspace{-4mm} $\phi_{1,2}$}
	\psfrag{J11}[l][][1.5]{\hspace{-5.5mm} $\V{S}_1^{(1)}$}
	\psfrag{J22}[l][][1.5]{\hspace{-3mm} $\V{J}_\text{e}(\V{p}_2^{(1)})$}
	\psfrag{J12}[l][][1.5]{\hspace{-3mm} $\V{J}_\text{e}(\V{p}_1^{(1)})$}
	\psfrag{J21}[l][][1.5]{\hspace{-5mm} $\V{S}_2^{(1)}$}
	\psfrag{v}[l][][1.5]{\hspace{-6mm} $\V{S}_{1,2}^{(1)}$}
	\psfrag{v1}[l][][1.5]{\hspace{0mm} $\V{S}_{1,2}^{(1)}$} \includegraphics[angle=0,width=0.46\linewidth,draft=false]{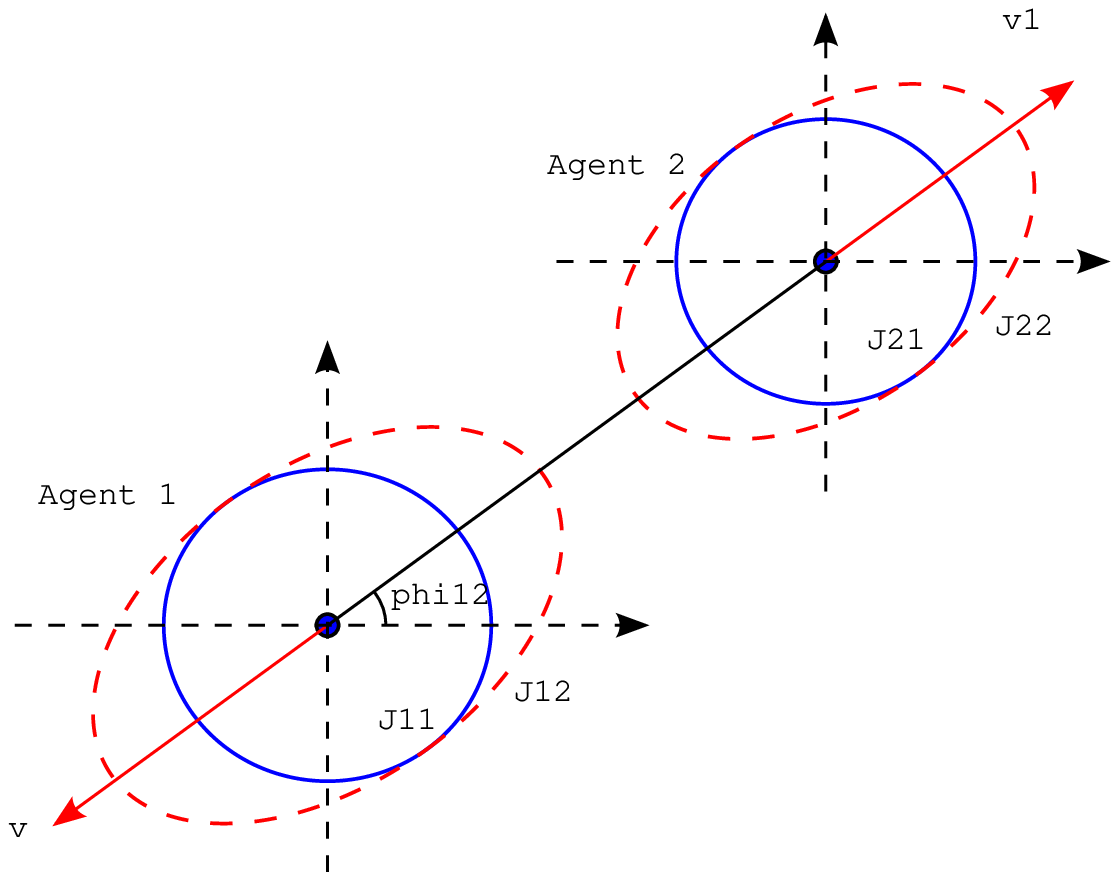}\label{fig:spatial_coop}}
	\vspace*{3.5mm}
	
	\caption{\label{fig:coop_spatem} Temporal and spatial cooperation: (a) Cooperation in time increases the navigation information both in the direction formed by the nodes positions in different times and in its orthogonal direction; (b) Cooperation in space increases the navigation information in the direction formed by the two nodes' positions.}
\end{figure}

\begin{figure}[t]
	\psfrag{agent1}[l][][0.9]{\hspace{-14mm} Agent 2}
	\psfrag{agent2}[l][][0.9]{\hspace{-14mm} Agent 1}
	\psfrag{time1}[l][][1]{\hspace{-6mm} Time 1}	
	\psfrag{time2}[l][][1]{\hspace{-6mm} Time 2}
	
	\psfrag{a11}[l][][0.9]{\hspace{-5mm} $\V{S}_2^{(1)}$}
	\psfrag{a21}[l][][0.9]{\hspace{-5mm} $\V{S}_1^{(1)}$}
	\psfrag{a12}[l][][0.9]{\hspace{-5mm} $\V{S}_2^{(2)}$}
	\psfrag{a22}[l][][0.9]{\hspace{-5mm} $\V{S}_1^{(2)}$}
	
	\psfrag{s11}[l][][0.9]{\hspace{-5mm} $\V{J}_\text{e}(\V{p}_2^{(1)})$}
	\psfrag{s21}[l][][0.9]{\hspace{-6mm} $\V{J}_\text{e}(\V{p}_1^{(1)})$}
	
	\psfrag{s12}[l][][0.9]{\hspace{-5mm} $\V{J}_\text{e}(\V{p}_2^{(2)})$}
	\psfrag{s22}[l][][0.9]{\hspace{-6mm} $\V{J}_\text{e}(\V{p}_1^{(2)})$}

	\psfrag{t1}[l][][0.9]{\hspace{-5mm} }
	\psfrag{t2}[l][][0.9]{\hspace{-6mm} }

	\centering
	{\includegraphics[angle=0,width=0.7\linewidth,draft=false]{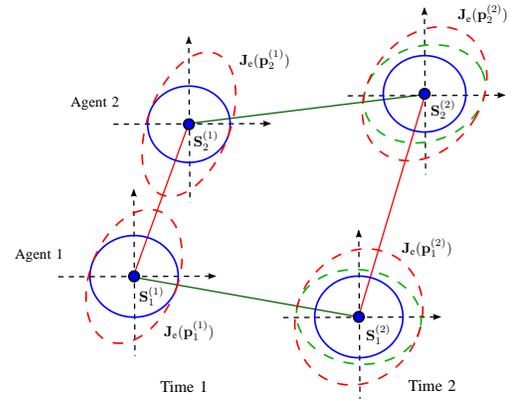}}

    \caption{Information evolution in time for network navigation with two agents in two consecutive time instants. The green and red ellipses denote the navigation information after temporal and spatial cooperation, respectively.
 \label{fig:spatem_total}}
\end{figure}

\begin{figure}[t]
	\centering
	\psfrag{Spatial cooperation }[c][][1.1]{Spatial cooperation}
	\psfrag{Temporal cooperation}[c][][1.1]{Temporal cooperation}
	\psfrag{Spatial and temporal cooperation}[c][][1.1]{\hspace{1mm}Temporal and spatial cooperation}
	\psfrag{TS}[c][][1.2]{Time Steps}
	\psfrag{AC}[c][][1.2]{Average SPEB (m$^2$)}
	{\includegraphics[angle=0,width=1\linewidth,draft=false]{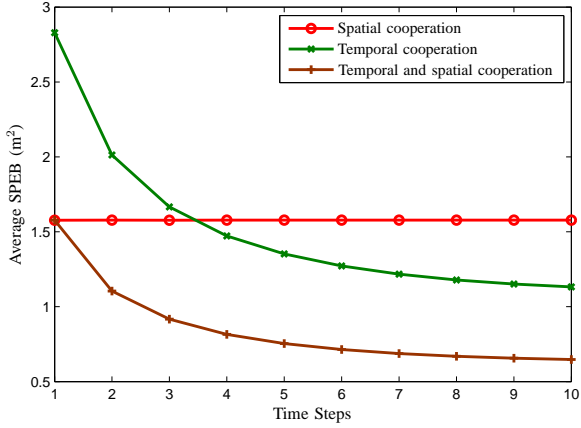}}
   
\caption{Average SPEB in cooperative navigation networks with respect to consecutive time steps. Agents randomly locate in an area of 20\thinspace m by 20\thinspace m and follow a Gaussian random walk, with $\lambda_{kk}^{(n)} = \nu_{kk}^{(n)} = \lambda_{kj}^{(n)} = 5\,\text{m}^{-2}$ and $\xi_{kk}^{(n)} = 0$. \label{fig:CRLB_numagent}}
\end{figure}

\begin{figure}[t]
	\centering

\psfrag{Spatial cooperation}[c][][1.1]{Spatial cooperation}
\psfrag{Temporal cooperation}[c][][1.1]{Temporal cooperation}
\psfrag{Spatial and temporal cooperation}[c][][1.1]{\hspace{1mm}Temporal and spatial cooperation}
\psfrag{NN}[c][][1.2]{Number of Nodes}
\psfrag{AC}[c][][1.2]{Average SPEB (m$^2$)}
{\includegraphics[angle=0,width=1\linewidth,draft=false]{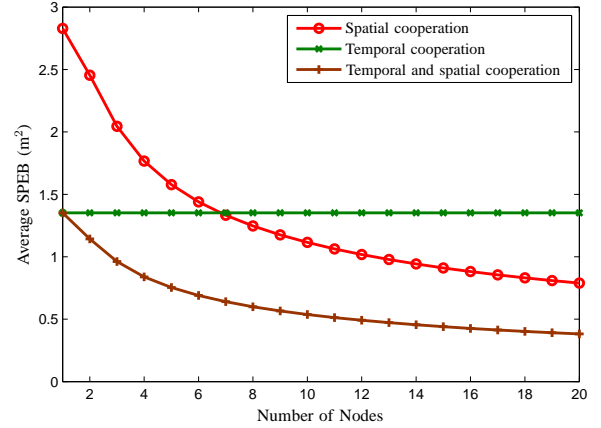}}

\caption{Average SPEB in cooperative navigation networks with respect to number of agents. Agents randomly locate in an area of 20\thinspace m by 20\thinspace m and follow a Gaussian random walk, with $\lambda_{kk}^{(n)} = \nu_{kk}^{(n)} = \lambda_{kj}^{(n)} = 5\,\text{m}^{-2}$ and $\xi_{kk}^{(n)} = 0$. \label{fig:CRLB_numagent1}}
\end{figure}

\begin{remrk}
The proof follows from a similar derivation of Theorem \ref{thm:EFIM_complete}. The EFIM for the agents' positions has the diagonally-striped structure as shown in (\ref{eq:example_1}), with each building block of size $2\Na \times 2\Na$. The EFIM $\V{S}^{(n)}$ characterizes the spatial cooperation at time $t_n$, and $\V{K}^{(n)}$ characterizes the temporal cooperation from time $t_{n-1}$ to $t_n$. Both of them consist of $2\times 2$ building blocks $\V{K}^{(n)}_k$ or $\V{S}_{kj}^{(n,n)}$ that correspond to either an intra- or inter-node measurement. When the amplitude and direction measurement of the velocity are independent, the corresponding $\V{K}^{(n)}_k$ can be decomposed into the direction of movement (from the amplitude measurement) and the orthogonal direction (from the direction measurement). On the other hand, the EFIM from the inter-node measurement is 1-D in the direction connecting the two nodes. 

Figures \ref{fig:temporal_coop}, \ref{fig:spatial_coop}, and \ref{fig:spatem_total} illustrate the contribution from temporal cooperation, spatial cooperation, and joint cooperation in terms of information ellipse, respectively.\footnote{The information ellipse characterized by $\{\V{p}\in \mathbb{R}^2: \V{p}^\text{T} \V{J}_\text{e}^{-1} \V{p} = 1\}$ for an EFIM $\V{J}_\text{e}\in\mathbb{S}_{++}^2$ \cite{SheWymWin:J10}.} Moreover, Fig.~\ref{fig:CRLB_numagent} and \ref{fig:CRLB_numagent1} show the squared position error bound (SPEB) \cite{SheWin:J10a} obtained from the EFIM for the agents' positions with different types of cooperation, where the contribution of the temporal and spatial cooperation increases with the time steps and the network size, respectively. These figures also show the significant performance improvement that can be achieved by joint temporal and spatial cooperation.
\end{remrk}

\subsection{Carry-over Information}

The diagonally-striped structure of (\ref{eq:example_1}) allows derivation of the EFIM $\J_\text{e}(\V{p}^{(T)})$ for the agents' positions $\V{p}^{(T)}$ at time instant $t_T$ recursively. For example, we can first apply the EFI to obtain the EFIM $\J_\text{e}({\V{p}}^{(2:T)})$. We next define the notion of \emph{carry-over} information, which characterizes the useful information transferred from one time instant to the next through temporal cooperation.

\begin{defi}[Carry-over information]
The carry-over information from $t_{n-1}$ to $t_n$ is defined to be the EFIM $\widetilde{\V{K}}^{(n)} \in \mathbb{R}^{D\times D}$ such that the EFIM $\J_\text{e}({\V{p}}^{(n:T)})$ can be written as (\ref{eq:EFIM_carry-over}) shown at the bottom of the page. \addtocounter{equation}{1}
\end{defi}

Note that the carry-over information $\widetilde{\V{K}}^{(n)}$ retains all the information from $t_{1}$ to $t_{n-1}$ for the EFIM $\J_\text{e}({\V{p}}^{(n:T)})$. In the following proposition, we show that such carry-over information always exists and derive its expression.
\begin{pro}\label{pro:general_case}
The carry-over information $\widetilde{\V{K}}^{(n)} \in \mathbb{S}_{++}^2$ always exists and is unique. It is given by
\begin{align}\label{eq:carryover_n}
	\widetilde{\V{K}}^{(n)} = {\V{K}}^{(n)} - {\V{K}}^{(n)} \left( {\V{S}}^{(n-1)}+ \widetilde{\V{K}}^{(n-1)} +{\V{K}}^{(n)} \right)^{-1} {\V{K}}^{(n)}
\end{align}
with $\widetilde{\V{K}}^{(1)}:=\V{0}$.
\end{pro}

\begin{remrk}
The proposition shows that the carry-over information for navigation can be obtained recursively at each time instant and used as prior knowledge of the agents' positions for the next time instant. In the following, we characterize the properties of this important information matrix to gain insights into temporal cooperation.

The spatial cooperation represented by ${\V{S}}^{(n-1)}$ in (\ref{eq:carryover_n}) is generally highly-coupled inference, i.e., the efficient estimates of the agents' positions after spatial cooperation are correlated \cite{MazSheWin:L11}. However, in distributed networks, the agents usually do not capture such correlation and only obtain individual (marginal) position distributions. Hence, after spatial cooperation at each time instant, the navigation accuracy limits of individual agents can be characterized by their own EFIMs, ignoring the correlation caused by spatial cooperation. We next consider a new EFIM consisting of the EFIMs for individual agents, given by
\begin{align}\label{eq:spatial_decouple}
	\widetilde{\V{S}}^{(n-1)} = \text{diag} \left\{\widetilde{\V{S}}^{(n-1)}_1, \widetilde{\V{S}}^{(n-1)}_2,\ldots, \widetilde{\V{S}}^{(n-1)}_{\Na} \right\}
\end{align}
where $\widetilde{\V{S}}^{(n-1)}_k = {\big\{} {\big[} ({\V{S}}^{(n-1)}+ \widetilde{\V{K}}^{(n-1)})^{-1}{\big]}_{\V{p}_k^{(n-1)}} {\big\}}^{-1}$ is the individual EFIM for agent $k$ at time $t_{n-1}$ after spatial cooperation.
\end{remrk}

\begin{pro}\label{pro:distributed_network}
In distributed networks, the carry-over information becomes
	\begin{align}\label{eq:carryover_2}
		\widetilde{\V{K}}^{(n)}  = \text{diag} {\bigg\{} \widetilde{\V{K}}^{(n)}_1, \widetilde{\V{K}}^{(n)}_2,\ldots, \widetilde{\V{K}}^{(n)}_\Na {\bigg\}}
	\end{align}
where $\widetilde{\V{K}}^{(n)}_k = \V{K}_{k}^{(n)} - \V{K}_{k}^{(n)} \left(\widetilde{\V{S}}^{(n-1)}_k  +\V{K}_{k}^{(n)}\right)^{-1} \V{K}_{k}^{(n)}$ is the carry-over information of agent $k$ from time $t_{n-1}$ to $t_n$. Moreover, (\ref{eq:carryover_n}) reduces to (\ref{eq:carryover_2}) for noncooperative navigation since the distributed condition is slack.
\end{pro}


\begin{remrk}
This proposition shows how the navigation information evolves in distributed navigation networks: at each time instant, each agent uses its own carry-over information as prior knowledge, updates its position distribution through spatial cooperation with its neighbors through a few iterations, and finally obtains the carry-over information for the next time instant based on its position distribution and temporal cooperation. This insight reveals that the complex joint cooperation can be decoupled into two simpler ones without loss of information. Such a finding can significantly reduce the complexity and facilitate the design of distributed network navigation algorithms. To visualize the information evolution, we develop a geometrical interpretation in the next section. 
\end{remrk}

\subsection{Geometrical Interpretation}

We now develop a geometrical interpretation of navigation information, particularly the carry-over information in (\ref{eq:carryover_2}), for 2-D distributed network navigation.\footnote{The geometrical interpretation for spatial cooperation can be found in \cite{SheWymWin:J10}. Here, the focus is on the discussion of the temporal cooperation.} Since the EFIM (\ref{eq:carryover_2}) is a block-diagonal matrix with block size of $2 \times 2$, we can focus on one agent and simplify the notation of the carry-over information as
\begin{align}\label{eq:carryover_geometrical}
	\widetilde{\V{K}} = \V{K} - \V{K} \left(\V{S} +\V{K}\right)^{-1} \V{K}
\end{align}
where $\V{K} := \V{C} + \V{D} = \lambda \, \R(\psi) + \nu \, \R(\psi+\pi/2)$ and $\V{S} := \mu \, \R(\beta) + \eta \, \R(\beta+\pi/2)$ by eigen-decomposition for matrices in $\mathbb{S}_{++}^2$ \cite{HorJoh:B85}. 
We then simplify the expression of $\widetilde{\V{K}}$ in two different ways,  providing different insights into the carry-over information for navigation. 
\begin{pro}\label{pro:first_interpretation}
The carry-over navigation information in (\ref{eq:carryover_geometrical}) can be written as
\begin{align}\label{eq:weighted_2D}
	\widetilde{\V{K}} = \frac{|\V{K}|}{|\V{S}+\V{K}|}\,\V{S} + \frac{|\V{S}|}{|\V{S}+\V{K}|}\,\V{K} \,.
\end{align}
\end{pro}

\begin{IEEEproof}
(Outline) The proof uses the fact that $(\V{K}+\V{S})^{-1} = \left[ {\text{adj}(\V{K})+\text{adj}(\V{S})}\right]/{|\V{K}+\V{S}|}$ for $\V{K},\V{S} \in \mathbb{S}_{++}^2$. 
\end{IEEEproof}

\begin{remrk}
The proposition shows that the carry-over information can be written as a weighted sum of the EFIM from temporal cooperation $\V{K}$ and that from spatial cooperation $\V{S}$. The intuition is that the carry-over contribution depends on both the original position information (after spatial cooperation) and the information obtained from the intra-node measurements. We next examine some special cases of the result given by (\ref{eq:weighted_2D}):
\begin{itemize}
	\item When the direction measurement of velocity is not available, i.e., $\V{K} = \V{C}$,  we have $|\V{K}| = 0$ and hence $\widetilde{\V{K}} =  {|\V{S}|}/{|\V{C}+\V{S}|}\cdot\V{C}$. In this case, the carry-over information is 1-D along the direction of the movement, and the weight is related to the directional position uncertainty of the agent at the previous time instant.\footnote{Detailed discussion about directional position uncertainty can be found in \cite{SheWymWin:J10}.}
	\item The carry-over information satisfies $\widetilde{\V{K}} \preceq \V{K}$ because the navigation information from the intra-node measurement cannot be fully utilized due to the uncertainty of the previous position characterized by $\V{S}$. The equality is achieved when $\V{S}$ goes to $\text{diag}\{\infty,\infty\}$, in which case such navigation information can be fully utilized.
	\item When $\beta = \psi$ or $\beta = \psi + \pi/2$, the carry-over information in the two orthogonal directions is decoupled, and can be calculated as
	\begin{align*}
		\widetilde{\V{K}} = \begin{cases}
		\frac{\lambda \mu}{\lambda+\mu} \,\R(\psi) + \frac{\eta \nu}{\eta+\nu} \,\R(\psi + \pi/2) \,, &  \beta = \psi \,, \\
		\frac{\lambda \eta}{\lambda+\eta} \,\R(\psi + \pi/2) + \frac{\eta \nu}{\eta+\nu} \,\R(\psi) \,, &  \beta = \psi + \pi/2 \,.
		\end{cases}
	\end{align*}
\end{itemize}
\end{remrk}

The result (\ref{eq:weighted_2D}) relies on a property of $\mathbb{S}_{++}^2$, which cannot be generalized to 3-D cases. We next show another geometrical interpretation of the carry-over information, which is also applicable to 3-D cases.

\begin{pro}\label{pro:second_interpretation}
The carry-over navigation information can be written as
\begin{align}\label{eq:carryover_second_interpretation}
	\widetilde{\V{K}} =  \zeta_1 \, \V{C} + \zeta_2 \, \V{D} + \kappa \, \R(\psi,\psi+\pi/2) 
\end{align}
where 
\begin{align*}
	\zeta_1 & = {\big(} {1+\lambda \,\V{u}_{\psi}^\dagger(\V{S} + \V{D})^{-1} \V{u}_{\psi}} {\big)}^{-1} \\
	\zeta_2 & = {\big(}{1+\nu \, {\V{u}}_{\psi}^{\perp\,\dagger}(\V{S} + \V{C})^{-1} {\V{u}}_{\psi}^\perp}{\big)}^{-1} \\
	\kappa & = -2 \,\lambda \nu \cdot \V{u}_{\psi}^\dagger (\V{S}+\V{C}+\V{D})^{-1} {\V{u}}_{\psi}^\perp \,.
\end{align*}
In particular, $\zeta_1 = {|\V{S}+\V{D}|}/{|\V{S}+\V{D}+\V{C}|}$, $\zeta_2 = {|\V{S}+\V{C}|}/{|\V{S}+\V{C}+\V{D}|}$, and $\kappa = {\lambda \nu \, (\eta-\mu) \sin(2(\psi-\beta)) }/{|\V{S}+\V{C}+\V{D}|}$ for 2-D cases.
\end{pro}


\begin{remrk}
Proposition \ref{pro:second_interpretation} shows that the carry-over information can be represented as a sum of three terms as shown in
(\ref{eq:carryover_second_interpretation}). The first two terms are weighted information of $\V{C}$ and $\V{D}$ from temporal cooperation, respectively, where the weights depend on the directional position uncertainty of the agent at the previous time instant.
The third term characterizes the coupling due to misalignment of eigenvectors corresponding to $\V{K}$ and $\V{S}$.

For 2-D cases, the eigenvalues and eigenvectors of $\R(\psi,\psi+\pi/2)$ are $\left(\frac{1}{2},\V{u}_{\psi+\pi/4} \right)$ and $\left(-\frac{1}{2},{\V{u}}_{\psi-\pi/4}\right)$. Hence, the coupling information increases the original EFIM in direction of $\psi - \pi/4$ and decreases with the same amount in the direction of $\psi + \pi/4$, if $\psi - \beta \in [0,\pi)$. This amount is proportional to the product of the $\sin(2(\psi-\beta))$ and the difference $\eta-\mu$. In particular, this coupling information vanishes when (i) $\mu = \eta$, (ii) $\psi-\beta = 0, \pi/2$, and (iii) either $\lambda$ or $\nu$ equals 0. The first two cases translate to the alignment of eigenvectors corresponding to $\V{K}$ and $\V{S}$, and in the last case the navigation information from temporal cooperation degenerates to 1-D.
\end{remrk}

\begin{figure*}
	[!b] \vspace*{4pt} 
	\hrulefill
	\normalsize%
	\setcounter{MYtempeqncnt}{\value{equation}} 
	\setcounter{equation}{27}
	\begin{align}\label{eq:K_n_m}
		\V{K}_k^{(n,m)} 
		& =  \begin{cases}
		\B\Phi_{\V{x}_{k}^{(n)}, \V{x}_{k}^{(n)}}\left(f(\B\eta_{k}^{(n+1)}|\V{x}_{k}^{(n)},\B\eta_{k}^{(n)}) \cdot f(\V{z}_{kk}^{(n)}|\V{x}_{k}^{(n)},\B\eta_{k}^{(n)})\right)  - \sum\limits_{l=n}^{T} \widetilde{\V{D}}^{(n,l)}_k \left[\widetilde{\V{B}}^{(l)}_k\right]^{-1} \widetilde{\V{D}}^{(n,l)\;\dagger}_k \,,  & m = n \\
		- \sum\limits_{l=m}^{T} \widetilde{\V{D}}^{(n,l)}_k \left[\widetilde{\V{B}}^{(l)}_k\right]^{-1} \widetilde{\V{D}}^{(m,l)\;\dagger}_k \,,  & n < m \leq T 
	\end{cases}
	\end{align}
	\vspace*{0pt}
	\hrulefill
	\begin{align}\label{eq:Gkj_n_m}
		\V{S}_{kj}^{(n,m)} 
		& = \begin{cases}
			\B\Phi_{\V{x}_{k}^{(n)}, \V{x}_{k}^{(n)}}\left( f(\B\kappa_{kj}^{(n+1)}|\V{x}_{kj}^{(n)},\B\kappa_{kj}^{(n)}) \cdot f(\V{z}_{kj}^{(n)}|\V{x}_{kj}^{(n)},\B\kappa_{kj}^{(n)})\right) - \sum\limits_{l=n}^T \widetilde{\V{E}}^{(n,l)}_{kj} \left[\widetilde{\V{C}}_{kj}^{(l)}\right]^{-1} \widetilde{\V{E}}^{(n,l)\,\dagger}_{kj}\\
		\quad + \,\B\Phi_{\V{x}_{k}^{(n)}, \V{x}_{k}^{(n)}}\left( f(\B\kappa_{jk}^{(n+1)}|\V{x}_{jk}^{(n)},\B\kappa_{jk}^{(n)}) \cdot f(\V{z}_{jk}^{(n)}|\V{x}_{jk}^{(n)},\B\kappa_{jk}^{(n)})\right) 
		-  \sum\limits_{l=n}^T \widetilde{\V{E}}^{(n,l)}_{jk} \left[\widetilde{\V{C}}_{jk}^{(l)}\right]^{-1} \widetilde{\V{E}}^{(n,l)\,\dagger}_{jk}
		\,, & \hspace{0mm} m=n \\
			- \sum\limits_{l=m}^T \left[ \widetilde{\V{E}}^{(n,l)}_{kj} \left[\widetilde{\V{C}}_{kj}^{(l)}\right]^{-1} \widetilde{\V{E}}^{(m,l)\,\dagger}_{kj} + \widetilde{\V{E}}^{(n,l)}_{jk} \left[\widetilde{\V{C}}_{jk}^{(l)}\right]^{-1} \widetilde{\V{E}}^{(m,l)\,\dagger}_{jk} \right]  \,,& \hspace{0mm} n< m \leq T
		\end{cases}
	\end{align}
    \setcounter{equation}{\value{MYtempeqncnt}} \vspace*{-6pt} 
\end{figure*}

\section{Conclusion}\label{sec:Conc}

In this paper, we established a general framework for cooperative network navigation to determine the fundamental limits of navigation accuracy. We applied the EFI analysis to derive the navigation information for both Bayesian and deterministic formulations of the network navigation problem. We showed that such information can be written as the sum of three parts that correspond to the mobility model, temporal cooperation, and spatial cooperation. Moreover, each part can be further decomposed into basic building blocks associated with each measurement and prior knowledge. We also developed a geometrical interpretation of the navigation information as well as its evolution in time, yielding important insights into the essence of network navigation. Our results not only provide performance benchmarks for cooperative navigation networks, but also guide network design and operation under performance/complexity trade-off.


\section{Acknowledgments}\label{Sec:Achn}
The authors would like to thank A.~Conti and R.~Cohen for their valuable suggestions and careful reading of the manuscript.


\appendices

\section{Proof of Theorem \ref{thm:EFIM_complete}} \label{apd:proof_EFIM_complete}

\begin{IEEEproof}
(Outline) 
The derivation of the EFIM for $\V{x}$ can be done in two steps: (i) identify the structure of the original FIM $\J(\B\theta)$, and (ii) apply the EFI analysis to obtain the EFIM. 

For temporal cooperation, we eliminate $\B\eta_{k}^{(n)}$ in time sequence. The matrix $\widetilde{\V{B}}^{(n)}_k$ denotes the EFIM for $\B\eta_k^{(n)}$ when $\B\eta_k^{(1:n-1)}$ are eliminated by the EFI process, given by
\begin{align*}
	\widetilde{\V{B}}^{(n)}_k & = 
	\begin{cases} \V{B}^{(1,1)}_k \,, & \hspace{-8mm} n=1 \\ 
	\V{B}^{(n,n)}_k - \V{B}^{(n-1,n)\;\dagger}_k \left[ \widetilde{\V{B}}^{(n-1,n-1)}_k\right]^{-1} \V{B}^{(n-1,n)}_k \,, \\ & \hspace{-8mm} n > 1
	\end{cases}
\end{align*}
in which
\begin{align*}
	\V{B}_k^{(n,m)} & = 
	\begin{cases}
		\B\Phi_{\B\eta_{k}^{(n)}, \B\eta_{k}^{(n)}}{\big(} f(\B\eta_{k}^{(n)}|\V{x}_{k}^{(n-1)},\B\eta_{k}^{(n-1)}) \, \\[2mm] \qquad \cdot f(\B\eta_{k}^{(n+1)}|\V{x}_{k}^{(n)},\B\eta_{k}^{(n)}) \, f(\V{z}_{kk}^{(n)}|\V{x}_{k}^{(n)},\B\eta_{k}^{(n)}){\big)}\,, \\ & \hspace{-18mm} m=n \\[1mm]
		\B\Phi_{\B\eta_{k}^{(n)}, \B\eta_{k}^{(n+1)}}{\big(} f(\B\eta_{k}^{(n+1)}|\V{x}_{k}^{(n)},\B\eta_{k}^{(n)}) {\big)}\,, \\& \hspace{-18mm} m=n+1 \,;
	\end{cases}
\end{align*}
and $\widetilde{\V{D}}^{(n-l,n+1)}_k$ denotes the cross-information for $\V{x}_k^{(n-l)}$ and $\B\eta_k^{(n+1)}$ when $\B\eta_k^{(1:n-1)}$ are eliminated by the EFI process, given by
\begin{align*}
	\widetilde{\V{D}}^{(n)}_k & = \V{D}^{(n,n)}_k \\
	\widetilde{\V{D}}^{(n-l,n+1)}_k & = 
	\begin{cases}
		{\V{D}}^{(n,n+1)}_k - \V{D}^{(n,n)}_k \left[\widetilde{\V{B}}^{(n,n)}_k\right]^{-1} \V{B}^{(n,n+1)}_k \,, \\& \hspace{-22mm} l=0  \\
		- \widetilde{\V{D}}^{(n-l,n)}_k \left[\widetilde{\V{B}}^{(n,n)}_k\right]^{-1} \V{B}^{(n,n+1)}_k \,, \\ & \hspace{-22mm} 1\leq l\leq n-1
	\end{cases}
\end{align*}
in which
\begin{align*}
	\V{D}_k^{(n,m)} 
	& = \begin{cases} \B\Phi_{\V{x}_{k}^{(n)}, \B\eta_{k}^{(n)}} {\big(} f(\B\eta_{k}^{(n+1)}|\V{x}_{k}^{(n)},\B\eta_{k}^{(n)}) \\[2mm] \qquad \qquad \quad \cdot f(\V{z}_{kk}^{(n)}|\V{x}_{k}^{(n)},\B\eta_{k}^{(n)}){\big)}\,, & \hspace{-4mm} m=n \\[1mm]
	\B\Phi_{\V{x}_{k}^{(n)}, \B\eta_{k}^{(n+1)}} {\big(} f(\B\eta_{k}^{(n+1)}|\V{x}_{k}^{(n)},\B\eta_{k}^{(n)}) {\big)}\,, \\ & \hspace{-4mm} m=n+1 \,;
	\end{cases}
\end{align*}

For spatial cooperation, we eliminate $\B\kappa_{kj}^{(n)}$ in time sequence. The matrix $\widetilde{\V{C}}_{kj}^{(n)}$ denotes the EFIM for $\B\kappa_{kj}^{(n)}$ when $\B\kappa_{kj}^{(1:n-1)}$ are eliminated by the EFI process, given by
\begin{align*}
	\widetilde{\V{C}}_{kj}^{(n)} & = \begin{cases} 
	\V{C}_k^{(1,1)} \,, & \hspace{-8mm} n = 1\\ 
	{\V{C}}_{kj}^{(n,n)} - {\V{C}}_{kj}^{(n-1,n)\;\dagger} \left[\widetilde{\V{C}}_{kj}^{(n-1,n-1)} \right]^{-1} {\V{C}}_{kj}^{(n-1,n)}\,, \\ & \hspace{-8mm} n > 1
\end{cases}
\end{align*}
in which
\begin{align*}
	\V{C}_{kj}^{(n,m)} & = \begin{cases}
	\B\Phi_{\B\kappa_{kj}^{(n)}, \B\kappa_{kj}^{(n)}} {\big(} f(\B\kappa_{kj}^{(n)}|\V{x}_{kj}^{(n-1)},\B\kappa_{kj}^{(n-1)})  \\[2mm] \qquad \cdot f(\B\kappa_{kj}^{(n+1)}|\V{x}_{kj}^{(n)},\B\kappa_{kj}^{(n)}) \, f(\V{z}_{kj}^{(n)}|\V{x}_{kj}^{(n)},\B\kappa_{kj}^{(n)}){\big)}\,, \\ &\hspace{-20mm} m = n\\[1mm]
	 \B\Phi_{\B\kappa_{kj}^{(n)}, \B\kappa_{kj}^{(n+1)}}{\big(} f(\B\kappa_{kj}^{(n+1)}|\V{x}_{kj}^{(n)},\B\kappa_{kj}^{(n)}) {\big)}\,, \\ &\hspace{-20mm} m =n+1\,;
	\end{cases}
\end{align*}
and $\widetilde{\V{E}}^{(n-l,n+1)}_{kj}$ denotes the cross-information for $\V{x}_k^{(n-l)}$ and $\B\kappa_{kj}^{(n+1)}$ when $\B\kappa_{kj}^{(1:n-1)}$ are eliminated by the EFI process, given by
\begin{align*}
	\widetilde{\V{E}}^{(n)}_{kj} & = \V{E}^{(n,n)}_{kj} \\
	\widetilde{\V{E}}^{(n-l,n+1)}_{kj} & = 
	\begin{cases}
		 \V{E}^{(n,n+1)}_{kj} - \V{E}^{(n,n)}_{kj}\left[\widetilde{\V{C}}_{kj}^{(n)}\right]^{-1} {\V{C}}_{kj}^{(n,n+1)} \,, \\ &\hspace{-17mm} l = 0\\
		- \widetilde{\V{E}}^{(n-l,n)}_{kj} \left[\widetilde{\V{C}}_{kj}^{(n)}\right]^{-1} {\V{C}}_{kj}^{(n,n+1)} \,, \\ &\hspace{-17mm}  1 \leq l \leq n-1
	\end{cases}
\end{align*}
in which
\begin{align*}
	\V{E}_{kj}^{(n,m)} & = \begin{cases}
	\B\Phi_{\V{x}_{k}^{(n)}, \B\kappa_{kj}^{(n)}} {\big(} f(\B\kappa_{kj}^{(n+1)}|\V{x}_{kj}^{(n)},\B\kappa_{kj}^{(n)}) \\[2mm] \qquad \qquad \quad \cdot f(\V{z}_{kj}^{(n)}|\V{x}_{kj}^{(n)},\B\kappa_{kj}^{(n)}){\big)}\,, & \hspace{-4mm} m = n\\[1mm]
	\B\Phi_{\V{x}_{k}^{(n)}, \B\kappa_{kj}^{(n+1)}}{\big(} f(\B\kappa_{kj}^{(n+1)}|\V{x}_{kj}^{(n)},\B\kappa_{kj}^{(n)}){\big)}\,, \\ & \hspace{-4mm} m = n+1\,.
	\end{cases}
\end{align*}
In particular, for notational convenience, we set $\widetilde{\V{E}}^{(n,m)}_{jk} = \V{0}$ for $j \in \NB$.

Combining all contributions from the EFI process, we have the EFIM for the positional state $\V{x}$ as in (\ref{eq:EFIM_complete}), where
\begin{align}\label{eq:p_n_m}
	\V{P}^{(n,m)}_k 
	& = \begin{cases}
	\B\Phi_{\V{x}_{k}^{(n)}, \V{x}_{k}^{(n)}}\left(f(\V{x}_{k}^{(n)}|\V{x}_{k}^{(n-1)}) \cdot f(\V{x}_{k}^{(n+1)}|\V{x}_{k}^{(n)}) \right)\,, \\ & \hspace{-16mm} m = 0 \\
	\B\Phi_{\V{x}_{k}^{(n)}, \V{x}_{k}^{(n+1)}}\left(f(\V{x}_{k}^{(n+1)}|\V{x}_{k}^{(n)}) \right)\,, & \hspace{-16mm} m = 1 \\
	\V{0} \,, & \hspace{-16mm} \text{otherwise}\,,
\end{cases}
\end{align}
and $\V{K}_k^{(n,m)}$ and $\V{S}_{kj}^{(n,m)}$ are given by (\ref{eq:K_n_m}) and (\ref{eq:Gkj_n_m}), respectively, shown at the bottom of the page.
\end{IEEEproof}

\bibliographystyle{IEEEtran}
\end{document}